\documentclass[journal]{IEEEtran}

\ifCLASSINFOpdf
\else
\fi
\usepackage{cite}
\usepackage{tikz}
\usepackage{amsmath}
\usepackage{filecontents}
\usepackage{xspace}
\usepackage{tabularx}
\usepackage{array} 
\usepackage{booktabs}
\usepackage{colortbl}
\usepackage[table]{xcolor}
\usepackage{multirow}
\usepackage{graphicx}
\usepackage{subfigure}
\usepackage{makecell}
\usepackage[most]{tcolorbox}
\usepackage{enumitem}
\usepackage{flushend}
\usepackage{setspace}
\usepackage{float}
\newfloat{pseudocode}{htbp}{lop}
\floatname{pseudocode}{Pseudocode}
\usepackage{tikz}
\newcommand{\blackcircnum}[1]{%
  \tikz[baseline=(char.base)]{
    \node[shape=circle,fill=black,inner sep=0.5pt, minimum size=0.5pt] (char)
    {\color{white}\bfseries #1};}}

\usepackage[colorlinks,
linkcolor=black,
anchorcolor=black,
citecolor=black,
urlcolor=black]{hyperref}

\usepackage{xcolor}
\usepackage{listings}
\usepackage{tcolorbox}
\tcbuselibrary{listings,skins,breakable}
\definecolor{CodeBG}{HTML}{FBFBFC}      
\definecolor{CodeFrame}{HTML}{D0D7DE}   
\definecolor{TitleBG}{HTML}{5E6A75}     
\definecolor{TitleFG}{HTML}{3B4350}     

\definecolor{KwBlue}{HTML}{0B64D8}   
\definecolor{StrRed}{HTML}{B31D28}  
\definecolor{CmtGreen}{HTML}{22863A}   
\definecolor{NameBrown}{HTML}{795E26} 
\definecolor{NumTeal}{HTML}{098658}
\definecolor{VarBlue}{RGB}{0,51,153}
\definecolor{KwPurple}{HTML}{AF00DB}
\definecolor{ModTeal}{HTML}{267F99}
\definecolor{ClassCyan}{HTML}{4EC9B0}
\definecolor{ConstBlue}{HTML}{4FC1FF}
\definecolor{StrOrange}{HTML}{CE9178}
\usepackage[T1]{fontenc}
\usepackage[scaled=0.85]{beramono}

\lstdefinestyle{vscodepython}{
  language=Python,
  basicstyle=\ttfamily\fontsize{10}{12}\selectfont\color{black},
  keywordstyle=\bfseries\color{KwBlue},
  stringstyle=\bfseries\color{StrRed},
  commentstyle=\itshape\bfseries\color{CmtGreen},
  numberstyle=\tiny\color{gray},
  numbers=left, numbersep=7pt,
  showstringspaces=false, keepspaces=true,
  breaklines=true, columns=fullflexible, tabsize=4,
  frame=none, framerule=0pt, rulesep=0pt,
  framexleftmargin=0pt, xleftmargin=1.1em,
  rulecolor=\color{white}, backgroundcolor=\color{white},
  aboveskip=0pt, belowskip=0pt, lineskip=-0.4pt,
  texcl=true,
  emph={build_xapp, configuration, io_contract, feature_spec, model_spec, KMeans, preprocess, norm, ewm, diff, detect, detector, run, inference_step, IsolationForest, postprocess, sanity_check, predict, RandomForest, format, validate, format_output, datetime, subscribe_to_kpis, send_policy_update, main, start, __init__, generate_synthetic_metrics, generate_policy, monitoring_loop, ric_interface, stop},
  emphstyle=\bfseries\color{NameBrown},
  emph={[2]{requirement, sub_config, contract, features, model, raw_kpm, state, data, st, x, y, decision, label, km, output, config, xapp, api, anomaly, out}
  },
  emphstyle={[2]\bfseries\color{VarBlue}},
  emph={[3]{True, False, None, self, __name__}},
  emphstyle={[3]\bfseries\color[HTML]{569CD6}},
  emph={[4]{import, as, if}
  },
  emphstyle={[4]\bfseries\color{KwPurple}},
  emph={[5]{numpy, np, seaborn, sns, sklearn, cluster, ensemble, utils, json, requests, datetime}
  },
  emphstyle={[5]\bfseries\color{ModTeal}},
  emph={[6]{ORANAnomalyXApp, AnomalyType, KPIMetrics, RICInterface, 
          AnomalyDetector, PolicyEngine, XAppAPIServer, AnomalyReporter, ConfigManager}},
  emphstyle={[6]\bfseries\color{ClassCyan}},
  emph={[7]{__main__}},
  emphstyle={[7]\color{StrOrange}}
}

\newtcblisting{prettycode}[2][]{%
  listing only, listing engine=listings,
  listing options={style=vscodepython},
  enhanced jigsaw, 
  colback=CodeBG, 
  colframe=black,   
  boxrule=0.8pt, arc=2mm,
  left=6pt, right=6pt, top=4pt, bottom=4pt,
  title={}, 
  before upper={%
    \par\centering{\bfseries\normalsize\color{TitleFG}#2}\par\vspace{0.25em}%
  },
  colbacktitle=CodeBG, coltitle=TitleFG, boxed title style={empty},
  frame code={\draw[black,line width=0.8pt,rounded corners=2mm]
    (frame.south west) rectangle (frame.north east);},
  interior style={fill=CodeBG},
  #1
}

\newtcolorbox{diffnote}{
  colback=yellow!6, colframe=yellow!40!black,
  boxrule=0.6pt, arc=2mm, left=6pt, right=6pt, top=4pt, bottom=4pt,
  fonttitle=\bfseries\normalsize, title=Key difference
}

\newtcolorbox{LLM}[1][]{
  colback=yellow!6, colframe=yellow!40!black,
  boxrule=0.6pt, arc=2mm, left=6pt, right=6pt, top=4pt, bottom=4pt,
  fonttitle=\bfseries\normalsize,
  title=#1
}

\usepackage{pifont}

\setenumerate[1]{itemsep=0pt,partopsep=0pt,parsep=\parskip,topsep=0pt}
\setitemize[1]{itemsep=0pt,partopsep=0pt,parsep=\parskip,topsep=0pt}
\setdescription{itemsep=0pt,partopsep=0pt,parsep=\parskip,topsep=0pt}

\newtcbtheorem[]{cmt}{Prompt}{
	colbacktitle = black!60!white, colframe = black!60!white,
	colback = black!5!white,
	fonttitle=\bfseries\footnotesize,
        fontupper = \footnotesize,
        right=3pt, left=3pt, top=2pt, bottom=2pt
}{t}

\newtcolorbox{templatebox}[1][]{
  colbacktitle=black!60!white,
  colframe=black!60!white,
  colback=black!5!white,
  fonttitle=\bfseries\footnotesize,
  fontupper=\footnotesize,
  title= #1,
  right=3pt, left=3pt, top=2pt, bottom=2pt
}
\usepackage{comment}

\newcommand{\update}[1]{\textcolor{black}{#1}}
\newcommand{\revise}[1]{\textcolor{red}{#1}} 
\newcommand{\etal}{\textit{et al.}}
\newcommand{\ie}{\textit{i}.\textit{e}.}
\newcommand{\eg}{\textit{e}.\textit{g}.}

\newcommand{\quotes}[1]{``#1''}

\newcommand{\projectName}{\textit{xAppGen}\xspace}
\usepackage{amssymb}
\begin{document}
%
\title{xAppGen: LLM-driven Natural Language \\ Programming for Agile xApp Development}
%
%
%
\author{Xin~Li,
        Shiming~Yu,
        Leming~Shen,
        Jianing~Zhang,
        Yuanqing~Zheng, 
        and~Yaxiong~Xie
 \thanks{Xin~Li, Shiming~Yu, Leming~Shen, Jianing~Zhang, and Yuanqing~Zheng are with the Department of Computing, The Hong Kong Polytechnic University, Hong Kong SAR, China (e-mail: cs-xin.li@connect.polyu.hk; shiming.yu@connect.polyu.hk; leming.shen@connect.polyu.hk; jianing98.zhang@connect.polyu.hk; csyqzheng@comp.polyu.edu.hk).}
 \thanks{Yaxiong~Xie is with the Department of Computer Science and Engineering, University at Buffalo, NY, USA (e-mail: yaxiongx@buffalo.edu).}
 \thanks{This work has been submitted to the IEEE for possible publication. Copyright may be transferred without notice, after which this version may no longer be accessible.}
}

%
%

\markboth{Submitted for review to IEEE Transactions on Mobile Computing}%
{xAppGen: LLM-driven Natural Language Programming for Agile xApp Development}
%



\maketitle

\begin{abstract}
Traditional RAN systems are closed and monolithic, stifling innovation. The openness and programmability enabled by Open Radio Access Network (O-RAN) are envisioned to revolutionize cellular networks with control-plane applications--xApps.
The development of xApps (typically by third-party developers), however, remains time-consuming and cumbersome, often requiring months of manual coding and integration, which hinders the roll-out of new functionalities in practice. To lower the barrier of xApp development for both developers and network operators, we present \projectName, the first LLM-driven natural language programming framework for agile xApps that automates the entire xApp development pipeline. In a nutshell, \projectName turns high-level user intents into swiftly deployable xApps within minutes, eliminating the need for manual coding or testing. To this end, \projectName builds a fully automated xApp generation pipeline, which integrates multiple functional modules (from user requirement elicitation, AI/ML function design and validation, to xApp synthesis and deployment). 
We design, implement, and comprehensively evaluate \projectName on representative xApp tasks.  
Results show \projectName-generated xApps can achieve similar or even better performance than the best known \update{manually developed} baselines. \projectName drastically accelerates the xApp development cycle (from user intent elicitation to roll-out), streamlining O-RAN innovation.
\end{abstract}


%
\IEEEpeerreviewmaketitle

\section{Introduction}
\label{sec:intro}
\IEEEPARstart{I}{n} recent years, Open Radio Access Network (O-RAN)~\cite{ORANAlliance,foukas2023taking,xing2023enabling,foukas2021concordia,lazarev2023resilient,foukas2025ranbooster,cheng2025integrated,chen2025sensing} has emerged as a promising paradigm for more intelligent and flexible cellular networks.
O-RAN disaggregates the traditional RAN into multiple modularized units, including the O-RAN Central Unit (O-CU), O-RAN Distributed Unit (O-DU), and O-RAN Radio Unit (O-RU), and standardizes the interfaces for interoperability. 
A key element of O-RAN is the RAN Intelligent Controllers (RICs), which act as the network's intelligence layer by hosting modular control-plane applications across different control timescales, including \textit{xApps} on the near-real-time (near-RT) RIC and \textit{rApps} on the non-real-time (non-RT) RIC. Specifically, xApps execute near-RT monitoring and control tasks at 10~ms to 1~s timescales, \eg, anomaly detection~\cite{sun2024spotlight,wen20246g,hong2025design} and traffic steering~\cite{lacava2023programmable,ntassah2023xapp} for self-optimizing networks.
This open and programmable architecture unleashes innovation by allowing third-party developers to introduce new RAN functionalities via xApps. However, it also creates an urgent need for \textbf{fast and flexible xApp development:} network operators must be able to develop and update xApps at the pace of evolving service demands.

\begin{figure}
    \centering
    \setlength{\abovecaptionskip}{2pt}
    \includegraphics[width=1\linewidth]{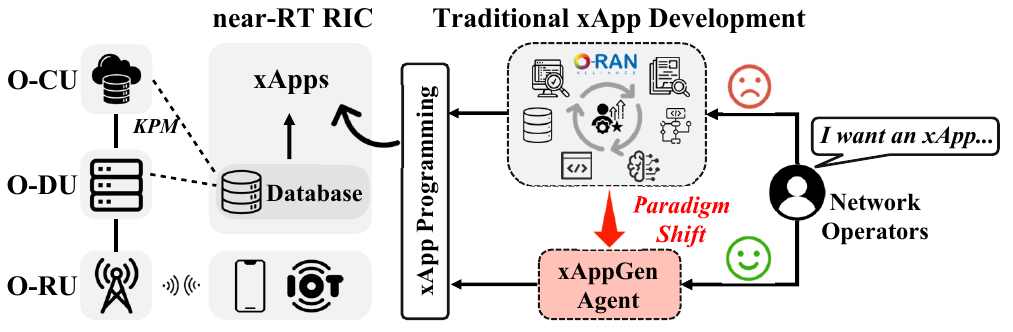}
    \caption{Conventional xApp development hinders O-RAN innovation. \projectName presents an automated agile xApp development framework.}
    \label{fig:idea}
    \vspace{-8pt}
\end{figure}

Developing a new xApp today, however, remains \textit{time-consuming and cumbersome}. It requires deep expertise in O-RAN architecture with complex interface specifications, protocols, and artificial intelligence/machine learning (AI/ML) algorithms for network control. Thus, third-party xApp developers often face tremendous challenges aligning with operators’ requirements, hindering the rollout of new functionalities. In practice, introducing a new service via xApps often requires substantial in-house development effort or reliance on external developers. In-house development demands substantial time and resources, while outsourcing raises concerns about data privacy and limited transparency. Fig.~\ref{fig:idea} illustrates the high-level gap between manual xApp development and the automated xApp generation paradigm enabled by \projectName.
Moreover, as networks scale and become more heterogeneous, operators may need to deploy hundreds of specialized xApps to cover diverse use cases~\cite{polese2023understanding,mungari2025ran,foukas2023taking,sapavath2023experimental,scalingi2024det}. This manual development paradigm has become a major bottleneck to O-RAN innovation, calling for a radically more efficient approach to xApp creation.

To address these challenges, we introduce \projectName, the first large language model (LLM)-driven natural language programming framework for agile xApp development. 
\update{Rather than directly prompting an LLM to generate a complete
xApp in one shot, \projectName follows a staged pipeline that refines operator intents, retrieves task- and platform-specific O-RAN information, and validates both the core logic and interface integration before deployment.}
Thus, a network operator can describe a desired network functionality or policy in natural language, while \projectName generates a corresponding deployable near-RT RIC xApp that satisfies the functional and interface requirements.

\update{This approach is inspired by recent advances in large language models (LLMs) and agentic AI~\cite{jiang2026survey,achiam2023gpt,fiandrino2023explora,aguzzi2025language}, which have exhibited the ability to interpret complex instructions and generate code from specifications. However, general-purpose LLMs (\eg, GPT-4~\cite{achiam2023gpt}) may generate hallucinated APIs or interface logic inconsistent with required specifications~\cite{chen2025towards}. 
Meanwhile, recent telecom- and O-RAN-specific LLMs have improved domain knowledge and code understanding~\cite{zou2025telecomgpt,gajjar2025oransight}, while other LLM-based systems have explored intent translation and specification-aware testing in O-RAN~\cite{al2026aura,ganiyu2025ai5gtest}. These studies demonstrate the value of incorporating O-RAN-specific knowledge and specification-aware validation into LLM-based O-RAN systems. However, these capabilities alone do not address the complete process of generating a deployable xApp from a high-level operator requirement. Such a process must identify the O-RAN information required by the requested function and target RIC/RAN platform, generate and validate the xApp function, synthesize the platform-specific interface layer, and validate the integrated xApp before deployment. Our key insight is to organize these requirements into a staged development pipeline that transforms high-level operator requirements into deployable near-RT RIC xApps.}

Realizing this vision requires addressing several O-RAN-specific challenges that make naive LLM-based xApp generation unreliable. \projectName addresses these challenges through a staged and domain-aware design.

\noindent\blackcircnum{1} \textbf{Under-specified requirements:} 
\update{Network operators specify high-level requirements for desired network functions or policies, while xApp developers translate these requirements into xApp implementations involving O-RAN interfaces, APIs, and service models. A natural-language request may not include all of the implementation details required for xApp development and can also be incomplete or ambiguous, causing an LLM to misinterpret the intended function~\cite{jacobs2021hey}. To mitigate this issue, \projectName uses a \textit{requirement refinement and structuring} module that guides the user to specify key xApp details (\eg, objectives, input metrics, policies). This structured template helps capture the intent and provides the LLM with a clear specification to work from.}

\noindent\update{\blackcircnum{2} \textbf{Task- and platform-specific O-RAN knowledge requirements:}
The O-RAN information required to generate a specific xApp varies with both the requested function and the target RIC/RAN platform. For example, different xApps may rely on different E2 service models, Key Performance Metric (KPM) semantics, platform APIs/SDKs, reporting scopes, and control interfaces. Having general O-RAN domain knowledge alone does not determine which of these specifications and platform capabilities should be used for a particular xApp, while missing or mismatched information can lead to incorrect API usage or interface logic~\cite{chen2025towards}.
\projectName addresses this issue through an automated \textit{knowledge retrieval} module, which retrieves relevant O-RAN specifications, platform information, and xApp development examples based on the refined user requirements and target deployment, and provides them as context to the LLM. This in-context knowledge injection~\cite{dong2024survey} provides the subsequent function design and xApp synthesis stages with the information required for the requested task and platform.}

\noindent\blackcircnum{3} \update{\textbf{Joint generation of xApp function logic and platform integration:} 
\projectName separates xApp generation into task-specific function logic and platform integration. The function logic may be rule-based, optimization-based, or AI/ML-based depending on the requested monitoring or control task. Platform integration refers to the implementation required to connect the generated function to the target near-RT RIC/RAN environment, including RIC lifecycle and messaging, E2 subscription setup, KPM parsing, and control-message encoding according to the applicable service models~\cite{santos2025managing,feraudo2024xdevsm}. Generating the function logic and platform integration in a single LLM step is \textit{error-prone}, since the LLM may mix task-specific algorithm requirements with O-RAN interface, service-model, and runtime requirements, causing functional or integration failures. \projectName tackles this challenge with a \textit{two-stage generation and validation process}: it first synthesizes and validates the standalone algorithmic \textit{core function}, and then uses the validated core as context to integrate the platform-specific interface code. This staged approach separates the two generation tasks and enables \projectName to produce a fully formed xApp that implements the requested function, satisfies the required O-RAN interfaces and target-platform constraints, and can be executed on the RIC without manual fixes.}

\begin{figure}
    \centering
    \setlength{\abovecaptionskip}{0pt}
    \includegraphics[width=0.9\linewidth]{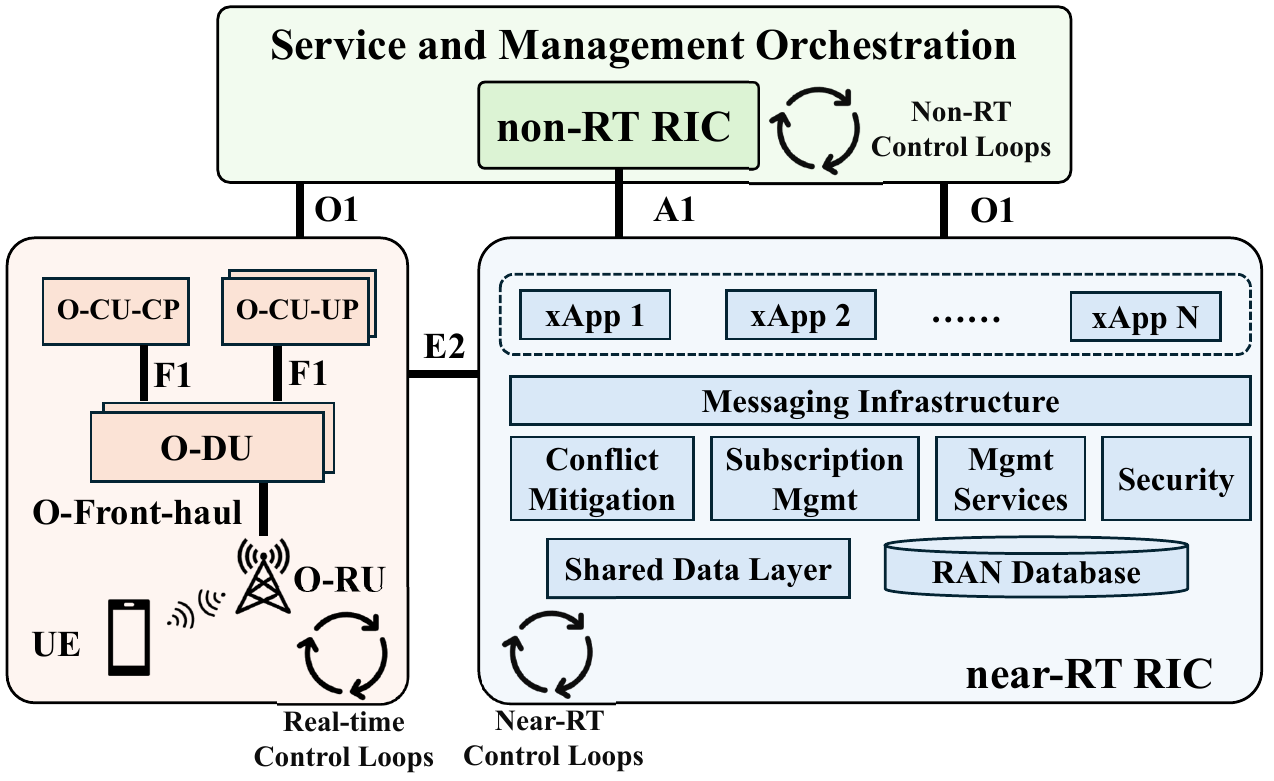}
    \caption{O-RAN architecture.}
    \label{fig:architecture}
    \vspace{-12pt}
\end{figure}

We build \projectName on a real-world 5G testbed using the open-source srsRAN stack~\cite{srsRAN}. We evaluate \projectName on multiple representative xApp use cases, including anomaly detection, interference classification, and slice scheduling, which cover various network monitoring and control scenarios. Results demonstrate that \projectName can generate functionally correct xApps that match or even exceed the effectiveness of the best \update{manually developed} ones. 
By eliminating most manual development effort, \projectName dramatically accelerates the xApp development pipeline --- a new xApp can be realized and rolled out in a matter of hours rather than weeks or months in the traditional paradigm.
In summary, this paper makes the following key contributions:

\begin{itemize}[leftmargin=8pt]
    \item We present \projectName, the first framework to automate xApp development using LLMs, transforming traditional manual processes into natural-language-driven workflows. This paradigm shift significantly reduces development effort and lowers the barrier for operators to create xApps.
    \update{\item We propose a suite of xApp-specific techniques to adapt LLMs into xApp development agents, integrating task- and platform-specific O-RAN knowledge retrieval with structured generation and validation for complex RAN interfaces and control logic. These techniques bridge high-level operator requirements and platform-specific xApp implementations with standard O-RAN interfaces and runtime validation.}
    \item We design, implement, and extensively evaluate \projectName on a live O-RAN testbed. Across diverse use cases, the \projectName-generated xApps demonstrate both strong functional performance and seamless deployment viability, achieving comparable or superior results to manually developed xApps while requiring minimal human effort.
\end{itemize}

\section{Background and Motivation}
\label{sec:related}
This section provides the background and motivation for \projectName. We first introduce the basic architecture of O-RAN and explain how its open interfaces enable programmable RAN control. We then describe the role of the near-RT RIC and xApps, including how xApps interact with RAN nodes through standardized interfaces. Finally, we discuss why existing xApp development remains difficult and motivate automated, domain-aware xApp generation.

\subsection{O-RAN v.s. Traditional RAN}
\begin{table}[t]
\centering
\setlength{\abovecaptionskip}{4pt}
\caption{Traditional RAN v.s. O-RAN }
\resizebox{0.45\textwidth}{!}{\begin{tabular}{c | c | c }
\specialrule{0.2em}{2.5pt}{2.5pt}
\textbf{Aspects} & \textbf{Traditional RAN} & \textbf{O-RAN} \\
\specialrule{0.2em}{2.5pt}{2.5pt}
\textbf{Architecture} & \makecell[c]{Monolithic, Integrated\\hardware/software} & \makecell[c]{Disaggregated into\\modular components} \\
\specialrule{0.1em}{2.5pt}{2.5pt}
\textbf{Interface} & Proprietary &  Standardized \\
\specialrule{0.1em}{2.5pt}{2.5pt}
\textbf{\makecell[c]{Vendor\\Ecosystem}} & \makecell[c]{Closed\\Single-vendor} & \makecell[c]{Open\\Multi-vendor} \\
\specialrule{0.1em}{2.5pt}{2.5pt}
\textbf{\makecell[c]{Control\\Logic}} & \makecell[c]{Embedded in\\software stack} & Realized through xApps \\
\specialrule{0.1em}{2.5pt}{2.5pt}
\textbf{\makecell[c]{Upgrade\\Flexibility}} & \makecell[c]{Constrained by\\vendor-specific dependencies} & \makecell[c]{High flexibility in\\upgrading components} \\
\specialrule{0.1em}{2.5pt}{2.5pt}
\textbf{\makecell[c]{Function\\Development}} & \makecell[c]{Black-box\\implementation} & \makecell[c]{Software-defined\\Programmable} \\
\specialrule{0.2em}{2.5pt}{2.5pt}
\end{tabular}}
\label{tab:RAN_ORAN}
\vspace{-12pt}
\end{table}

Traditional RAN architectures are built on vertically integrated hardware and software stacks provided by a single vendor. While this approach simplifies system integration, it leads to 
limited flexibility, increased cost, and vendor lock-in. Thus, launching new features often necessitates hardware upgrade by vendors, which is costly and time-consuming.

Unlike traditional RANs that rely on monolithic, vendor-locked solutions, O-RAN promotes interoperability by disaggregating RAN into multiple units with distinct functionalities. 
In~\figurename~\ref{fig:architecture}, the O-RU handles \update{Radio Frequency (RF)} signal processing and lower-PHY functions; the O-DU executes lower-layer protocol stack functions, including RLC, MAC, and higher-PHY functions; and the O-CU hosts higher-layer RAN protocol functions, including RRC and PDCP.
Meanwhile, to enable agile and intelligent control, O-RAN separates RAN control intelligence into the near-RT RIC~\cite{ORAN2025NRTRIC} and the non-RT RIC~\cite{ORAN2025NONRTRIC}. Operating at timescales ranging from 10 milliseconds to 1 second, the near-RT RIC enables dynamic control with xApps. Information exchange between the two RICs is supported via the A1 interface. 
By defining the E2 interface between the near-RT RIC and E2 Nodes~\cite{ORAN_E2GAP} and other open interfaces between the units, O-RAN supports the interoperability of units from different vendors and offers network operators more unit options.
\tablename~\ref{tab:RAN_ORAN} summarizes the key differences between traditional RAN and O-RAN.

\subsection{xApp Workflow}
\begin{figure}[t]
\setlength{\abovecaptionskip}{0pt}
    \centering
    \includegraphics[width=0.9\linewidth]{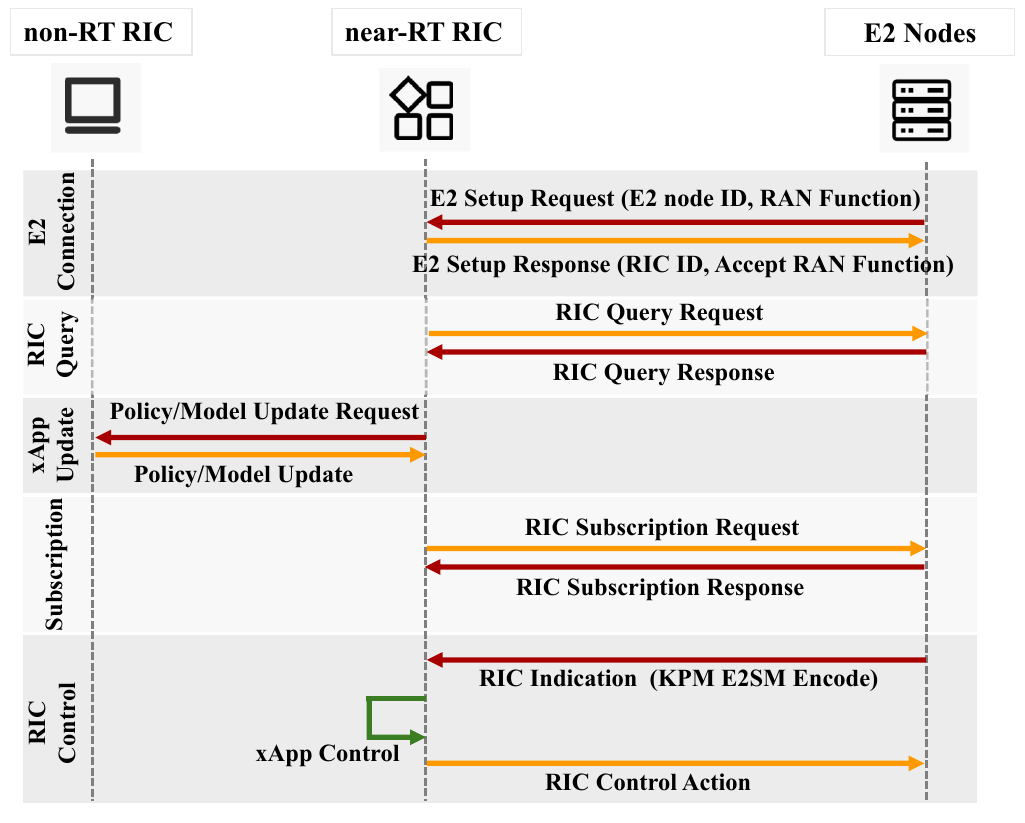}
    \caption{xApp Workflow.}
    \label{fig:xappworkflow}
    \vspace{-10pt}
\end{figure}

xApps enable fine-grained, near-RT control over O-RAN operations. These applications, hosted on the near-RT RIC, subscribe to specific telemetry streams (\eg, network KPMs), execute control logic, and transmit decisions to underlying RAN units via the E2 interface. 
\update{In this workflow, an E2 Node is a logical node that terminates the E2 interface and exposes RAN functions to the near-RT RIC. E2 Nodes include O-CU-CP, O-CU-UP, O-DU, and O-eNB.}
Each xApp follows a structured lifecycle including deployment, registration, optional RIC query, data subscription, decision execution, and control feedback. Specifically, upon deployment, an xApp registers its service capabilities and communication endpoints with the RIC. After E2 setup, the near-RT RIC may issue a RIC Query Request and receive a RIC Query Response to obtain RAN- or UE-related information from the E2 Node before or during subsequent subscription and control procedures. It then subscribes to measurements such as signal-to-noise ratio, handover events, or buffer occupancy, streamed in near-RT from E2 Nodes or other RAN nodes. The xApp processes the stream using control logic---typically rule-based policies or AI/ML models---and issues corresponding control actions, \eg, adjusting scheduling priorities or triggering handovers. This closed-loop interaction enables responsive network control. In parallel, the non-RT RIC can influence xApp behavior via the A1 interface, delivering high-level policy directives, model updates, or intent-driven objectives. \figurename~\ref{fig:xappworkflow} shows the typical workflow of an anomaly detection xApp.

\subsection{Current xApp Development Process}
\begin{figure}[t]
\setlength{\abovecaptionskip}{0pt}
    \centering
    \includegraphics[width=1\linewidth]{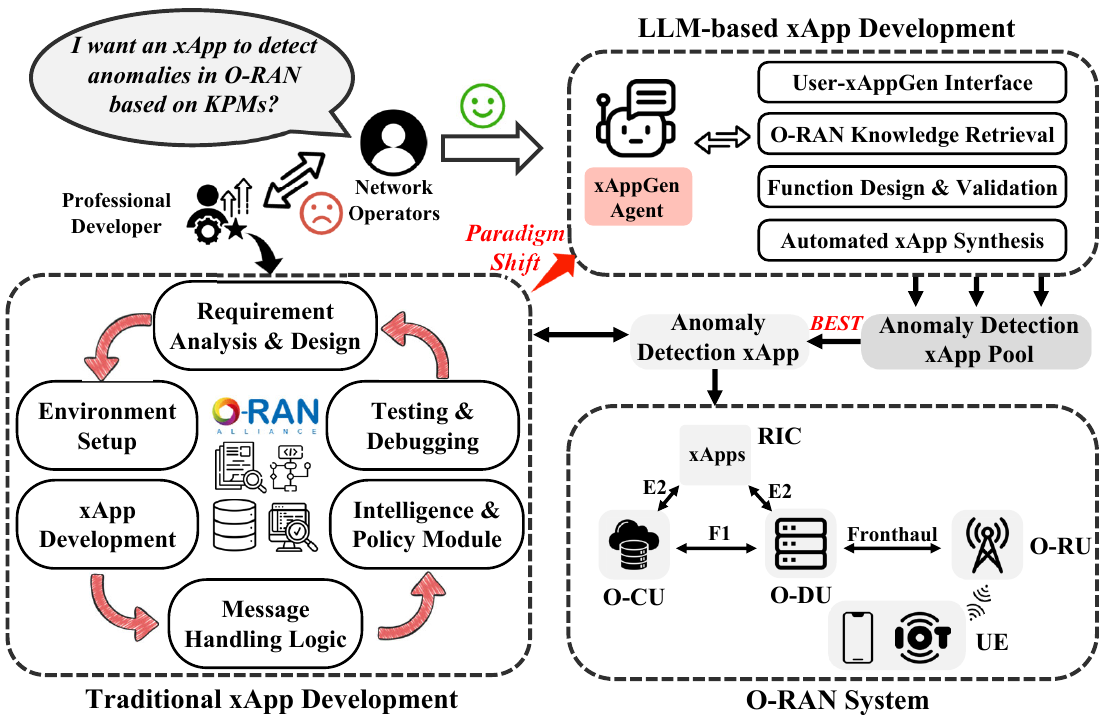}
    \caption{Traditional xApp Development in O-RAN v.s. Automated xApp Development.  }
    \label{fig:AutORAN_RAN}
    \vspace{-10pt}
\end{figure}

\update{Unlike proprietary monolithic RANs, where control functions are implemented within vendor-specific network equipment and access to these functions is limited by proprietary interfaces, O-RAN enables external xApps to program RAN behavior through open interfaces.}
This openness introduces a new development burden for xApp developers, who must navigate heterogeneous vendor units, constantly evolving specifications, RIC lifecycle management, O-RAN interfaces, service-model semantics, telemetry subscription, and control-message encoding.
Developing a deployable xApp therefore requires more than designing the control algorithm itself. 
\update{Developers must integrate the algorithm with the near-RT RIC runtime interfaces, configure E2 subscriptions, parse service-model-specific KPM reports, encode control messages according to the corresponding O-RAN service-model definitions, and validate the xApp against RIC lifecycle and latency constraints, all of which require deep domain expertise and familiarity with the O-RAN specifications (\figurename~\ref{fig:AutORAN_RAN}).}
As a result, practical xApp development is time-consuming (spanning weeks or months) even for experienced developers. For network operators, this not only delays innovation but also increases reliance on third-party developers, hindering agility in xApp development, raising concerns about user data, and increasing the deployment costs of new functionalities. These challenges highlight the pressing need for an agile xApp development paradigm.

\subsection{Towards Automated xApp Development}

\begin{figure*}[t]
\setlength{\abovecaptionskip}{0pt}
    \centering
    \includegraphics[width=1\linewidth]{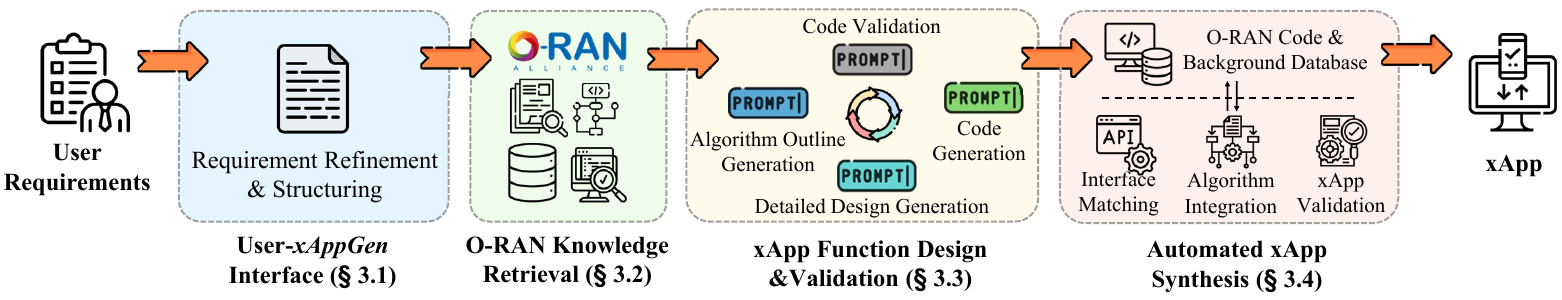}
    \caption{\projectName overview.}
    \label{fig:system_overview}
    \vspace{-10pt}
\end{figure*} 

Recent advances in LLMs (\eg, GPT-4o \cite{hurst2024gpt}, DeepSeek-Coder \cite{guo2024deepseek}) have substantially improved program synthesis and natural-language programming capabilities. By translating high-level user requirements into executable programs, LLMs offer an opportunity to reduce the effort required in traditional software development. This capability is particularly appealing for xApp development, where developers must bridge high-level network objectives and low-level RAN interfaces.

This motivates a natural question: can xApp development be shifted from manual function-level programming to intent-driven generation? \figurename~\ref{fig:AutORAN_RAN} illustrates this transition from conventional xApp development to LLM-based xApp generation. In the traditional workflow, xApp development requires repeated effort in requirement analysis, environment setup, implementation, message-handling logic, intelligence and policy integration, and testing/debugging. In contrast, xApp developers or network operators may express their needs in natural language, and an AI agent can translate these high-level requirements into executable xApps. This shift from function-level programming (by professional xApp developers) to intent-driven xApp generation (initiated by either developers or network operators) has the potential to significantly streamline xApp development~\cite{AIRANAlliance}.



\update{At the same time, recent studies have expanded the use of LLMs in telecom and O-RAN. Telecom-specific LLMs have been developed to improve domain knowledge and task performance~\cite{zou2025telecomgpt}, while other studies use LLMs to translate natural-language intents into O-RAN policies or management actions~\cite{al2026aura}. More recently, ORANSight-2.0~\cite{gajjar2025oransight} develops O-RAN-specific foundation models and demonstrates the benefit of domain-specific knowledge for O-RAN knowledge understanding and code-related tasks. AI5GTest~\cite{ganiyu2025ai5gtest} further demonstrates that 3GPP and O-RAN specifications can be incorporated into LLM-based systems for automated testing and validation of O-RAN components.}

\update{These advances provide important evidence for automated xApp development. ORANSight-2.0 shows that O-RAN-specific knowledge improves an LLM's ability to handle O-RAN tasks, while AI5GTest shows the importance of specification-aware reasoning and validation when LLMs interact with O-RAN systems. They support two key requirements of \projectName: the generation process should use O-RAN-specific information relevant to the target task, and the generated software should be validated against O-RAN and runtime requirements. However, automated xApp development requires these capabilities to be connected with additional steps, including operator requirement refinement, xApp function generation, platform-specific interface synthesis, deployment, and runtime execution.}

\update{Motivated by this gap, \projectName integrates task- and platform-specific O-RAN knowledge retrieval, xApp function generation and validation, platform-specific interface synthesis, deployment, and runtime validation into an automated development pipeline that transforms a high-level operator requirement into a deployable near-RT RIC xApp. In the following, we present the design of \projectName and its key technical components.}

\section{xAppGen Design}
\label{sec:model}
This section presents the design rationale and technical modules of \projectName. Automatic xApp generation requires more than translating a natural-language prompt into code. It must transform an underspecified network intent into a deployable, platform-specific near-RT RIC xApp. \update {To this end, \projectName addresses four main challenges: incomplete user intents, task- and platform-specific O-RAN knowledge requirements, O-RAN-aware core-function generation and validation, and platform-specific xApp synthesis under deployment and runtime constraints. Accordingly, \projectName is organized into four components, as shown in \figurename~\ref{fig:system_overview}. The User-\projectName Interface refines natural-language requests into structured xApp requirements; the O-RAN Knowledge Retrieval module retrieves O-RAN and platform information associated with the requested xApp and target deployment; the xApp Function Design and Validation module generates O-RAN-aware core functions and validates their algorithmic performance; and the Automated xApp Synthesis module performs interface matching, algorithm integration, and validation to produce deployable platform-specific xApps. This pipeline reduces manual xApp development effort and allows operators to specify new xApp functions at a higher level of abstraction.}

\subsection{User-xAppGen Interface}
\label{sec:user_interface}

Traditional xApp development requires mastering detailed O-RAN interfaces, data streams, and APIs, posing a high entry barrier. \projectName addresses this by providing a user-friendly interface where developers or operators specify desired functionalities in natural language. This design hides O-RAN complexity and enables non-experts to quickly develop and deploy xApps.
Specifically, users first specify task requirements, operational objectives, and high-level control policies in natural language as \quotes{Design a Python-based xApp to detect anomalies in O-RAN based on KPM measurements}. 
Such natural language specifications are intentionally designed to abstract away the details of implementation, allowing non-expert users to focus on expressing their intents. 


\noindent\textbf{Technical challenge: underspecified xApp intents.}
LLMs cannot generate high-quality code to meet user expectations when prompted with simple instructions~\cite{abbassi2025unveiling}. 
The main reason is that user requirements are often unstructured and underspecified, and lack sufficient details for correct design and implementation~\cite{jia2025automated,yang2026assessing,jacobs2021hey}. Therefore, the corresponding design choice is to refine natural-language requests into structured xApp requirements before code generation.
To tackle this challenge, we develop a \textit{requirement refinement and structuring} module with a user requirement template specifically tailored for xApp development.

\noindent\textbf{Requirement Refinement and Structuring.} 
\update{This module aims to progressively guide users to express structured requirements through multi-round interactions.} \projectName first parses the initial natural-language request using a lightweight intent-extraction module, which identifies key task components such as the functional objective (\eg, anomaly detection or traffic classification), expected input modality (\eg, structured telemetry or RF spectrograms), temporal resolution, and output format.
The operator may optionally specify the target RIC/RAN platform or other available deployment information. When no such information is provided, \projectName uses an O-RAN Software Community (OSC)-based reference profile~\cite{ORANSC,ORANSC_RICPLT,ORANSC_XAPPFRAME_PY} as the default generation target. The resulting xApp is generated for the selected reference profile and is not assumed to be portable across platforms.

If any fields are missing or under-specified, they are flagged as unresolved fields and trigger follow-up interactions. To handle ambiguous or incomplete input, \projectName can engage users in follow-up dialogues, automatically generating targeted questions to elicit additional details---such as the specific type of anomalies, relevant data sources, required measurement granularity, or deployment information needed to resolve an interface ambiguity. To determine what information needs to be clarified, \projectName maps the initial user input to a predefined \textit{User Requirement Template} corresponding to the intended xApp type (\eg, anomaly detection, interference classification, traffic classification). This template specifies essential fields such as the task objective, input modality, temporal resolution, and output format, ensuring that requirements are captured in a consistent and complete form. 

\subsection{O-RAN Knowledge Retrieval}
\label{sec:knowledge}
To bridge the gap between user intents specified in user requirement templates and detailed algorithm design and implementation, \projectName develops an \textit{O-RAN knowledge retrieval} module, which gathers relevant knowledge from diverse sources (\eg, O-RAN specifications) to facilitate accurate interpretation of user intents and xApp generation.


\noindent\update{\textbf{Technical challenge: selecting O-RAN knowledge for xApp generation.}}
Automatically generating a deployable xApp requires concrete O-RAN knowledge, including service-model definitions, KPM meanings, RIC platform information, SDK usage, and existing xApp development examples.
\update{The key challenge is to identify the subset of this information that applies to the current xApp requirement and deployment environment. For example, the generated xApp must select KPMs available at the required reporting scope, use service models supported by the target RIC/RAN platform, and invoke the corresponding platform APIs and control interfaces. Incorrect selections may result in unsupported KPMs or service models, or interface logic inconsistent with the required O-RAN specifications or target-platform APIs. Therefore, \projectName retrieves and selects task- and platform-specific O-RAN knowledge before algorithm design and xApp synthesis, and provides the selected information to the following generation stages.}

\noindent\textbf{Keyword Extraction.}
\update{\projectName first automatically identifies keywords from the standardized user requirements in the context of O-RAN.} This is important for retrieving highly relevant domain knowledge. Naive methods suffer the risk of missing essential contextual information or adding irrelevant noise \cite{zhu2025large}. For instance, extracting generic terms like \quotes{anomaly detection} without specifying the application domain may retrieve less relevant knowledge from other fields such as finance or healthcare. Conversely, extracting excessively fine-grained keywords may limit the flexibility and generalizability of the knowledge retrieval module. To address this challenge, \projectName adopts a structured keyword extraction strategy that decomposes each keyword into two semantic fields: the \textit{functional task} (\eg, \quotes{anomaly detection}, \quotes{traffic classification}) and \textit{target domain} (\eg, \quotes{in O-RAN}, \quotes{in near-RT RIC}, \quotes{based on KPMs}). \update{This two-part structure ensures that the extracted keywords reflect both the algorithmic intent and the deployment context.} As shown in Prompt 1, the system explicitly instructs LLMs to balance specificity and generality by extracting phrases with optimal granularity---for example, preferring \quotes{anomaly detection in O-RAN} over either \quotes{anomaly detection} (too general) or \quotes{anomaly detection in O-RAN based on past hour KPMs} (which is too specific and can be revised to \quotes{anomaly detection in O-RAN} and \quotes{KPMs in O-RAN}). This trade-off ensures that the retrieved domain knowledge is both highly relevant and widely applicable across similar xApp tasks.

\vspace{-3pt}
\begin{cmt}{Keyword Extraction}{}
\textbf{*User Problem*}\\
\textless user input\textgreater...\textless/user input\textgreater\\
\textbf{*Target*}\\
To understand the user problem, please identify key concepts that provide essential background knowledge.\\
\textbf{*Rules*}\\
Extract task-related and domain-related keyword phrases from the user problem. The keyword should contain two parts: the core task or function, and the domain context. Avoid generic or overly specific expressions.\\
\textbf{*Response Format*}\\
Term1, Term2, ...
\end{cmt}
\vspace{-3pt}



\noindent\textbf{Efficient Knowledge Storage and Retrieval.}
\projectName first checks the local knowledge base for relevant technical repositories, O-RAN specifications, service-model definitions, SDK documents, and open-source xApp development libraries. If the required information has already been collected, \projectName retrieves the corresponding documents from the local index. External web search is invoked only during initial knowledge-base construction or when the local repository lacks information for a new xApp requirement, target RIC platform, service model, or software version. To enhance efficiency and reliability, the search is scoped to a predefined list of selected authoritative sources (as in Prompt~2), ensuring that the retrieved content is both precise and relevant to the keyword and the development task.

The search results are then passed into the structuring and embedding pipeline for integration into local knowledge base. With various retrieved relevant information, \projectName converts them into a structured format (\eg, Markdown). Next, \projectName uses an embedding model to encode available knowledge items into a dense vector space (\ie, embeddings), enabling semantic-level retrieval beyond simple keyword matching. Finally, the embeddings are organized and stored in a local knowledge base, categorized according to content types such as O-RAN background and specifications, algorithm principles, performance optimization methods, and coding patterns.
During the entire code generation process, \projectName proactively retrieves relevant information from the knowledge base for reference. For instance, when a user requests for \quotes{anomaly detection in O-RAN}, the knowledge base provides relevant information about common KPMs for the functionality in the O-RAN literature, best practices for AI/ML model design for xApps, and a variety of performance evaluation metrics.
\begin{cmt}{Knowledge Search}{}
\textbf{*User Problem*}\\
<user input>\{...\}</user input>\\
\textbf{*Target*}\\
To enrich domain knowledge, please search for the fundamental definitions or background of given terms \{...\}.\\
\textbf{*Rules*}\\
Prioritize Wikipedia or official sites. Exclude implementation details. Filter out irrelevant content.\\
\textbf{*Response Format*}\\
URL1, URL2, ...
\end{cmt}

\subsection{xApp Function Design and Validation}
\label{sec:function_design}
This module focuses on generating executable code for the xApp core function with automated validation. 
Given the structured user requirement and the retrieved O-RAN knowledge, \projectName first translates the target O-RAN task into task-specific algorithmic logic, including telemetry feature selection, data preprocessing, model construction, decision logic, and evaluation metrics.

\noindent\textbf{Technical challenge: O-RAN-aware core-function generation.}
The core function of an xApp must be consistent with the O-RAN task scenario and achieve reliable algorithmic performance. However, directly generating such a function is challenging because different xApp tasks rely on different telemetry modalities, KPM semantics, feature-processing steps, model choices, and optimization objectives. A generic algorithm may be executable, but it may use irrelevant inputs, miss important O-RAN-specific features, or perform poorly on the target task. Therefore, the corresponding design choice is to decompose the core-function design into explicit algorithmic steps and validate the generated function with task-relevant datasets.

This module is designed to produce high-quality, optimized algorithms tailored to fulfill user requirements, which is accomplished through a multi-stage prompting framework guided by Chain-of-Thought (CoT) reasoning strategies~\cite{wei2022chain}. 
In particular, CoT reasoning is used to decompose the xApp core-function design into intermediate steps, including feature selection, preprocessing, model design, prediction or control logic, and evaluation. This decomposition helps \projectName check whether the generated algorithm matches the O-RAN task context and provides checkpoints for iterative correction and performance improvement.
The divide-and-conquer process has three key stages: algorithm outline generation, detailed design generation, and code generation with validation.

\begin{cmt}{Algorithm Outline Generation}{}
\textbf{*User Problem*}\\
\textless user input\textgreater\{...\}\textless/user input\textgreater\\
\textbf{*Target*}\\
To address the user problem effectively, please provide an algorithm outline step by step to solve the user problem based on the background information.\\
\textbf{*Rules*}\\
Provide a step-by-step algorithm design. Each step should have a clear goal and describe what actions will be performed. Focus on input/output relationship between steps.\\
\textbf{*Response Format*}\\
Step 1: [Title] ... ; Step 2: [Title] ... ; ...
\end{cmt}

\noindent\textbf{Algorithm Outline Generation.}
Considering the high complexity of xApp algorithms, \projectName is instructed to first generate a high-level outline of the target algorithm, decomposing the solution into a sequence of logical steps or functional modules. To ensure the relevance and clarity of the output, \projectName employs a specially designed Prompt 3 that reiterates the user problem description, specifies the generation target, and enforces strict quality assurance rules for consistency and task alignment. For example, when a user specifies an anomaly detection task based on O-RAN KPMs, \projectName generates an outline that begins with dataset loading and preprocessing, followed by feature selection, model construction, prediction logic, and evaluation using provided metrics (\eg, accuracy, F1 score). Each step is designed to ensure input-output consistency, robustness to missing data, and compatibility with a standalone execution script.

\noindent\textbf{Detailed Design Generation.}
Subsequently, \projectName is further prompted to expand on each step in the outline by specifying concrete operations, data processing methods, feature selection strategies, and decision criteria. 
For example, in the context of anomaly detection, the LLM may recommend selecting E2SM-KPM measurements~\cite{ORAN_E2SM_KPM}, such as PRB utilization, user throughput, or handover rates, as critical input features for training the detection model.

\begin{cmt}{Code Validation and Repair}{}
\textbf{*User Problem*}\\
<user input>\{...\}</user input>\\
\textbf{*Target*}\\
To ensure the generated code is executable and robust, please analyze the provided output logs \{...\}, identify the problem, and modify the code to fix the errors if the compiler or interpreter cannot successfully run the code.\\
\textbf{*Rules*}\\
The following case is not allowed:\\
\# ... (other functions remain unchanged)\\
\# ... (same as before)
\end{cmt}
\noindent\textbf{Code Generation with Validation.}
Once the detailed design is completed, \projectName triggers the \textit{Code Generation} module to translate each design specification into executable code segments. Then, \projectName constructively integrates them into a comprehensive program (\ie, the algorithm part of an xApp). In practice, however, the input data is collected from different O-RAN interfaces and protocols, exhibiting significant variability and lacking ground truth labels for model training. This limitation hinders \projectName from automatically verifying the correctness of the algorithm and even improving its performance iteratively. To tackle this challenge, we propose to instruct \projectName to first download public datasets corresponding to the user problem (\eg, SpotLight \cite{sun2024spotlight} for anomaly detection). 
The dataset is then input into the generated algorithm for evaluation, and the error logs are recorded and fed back to \projectName for iterative refinement, with the designed prompt shown in Prompt~4.
Such a self-correcting loop produces stable and functional code with minimal human intervention. Additionally, \projectName could generate multiple algorithm variants for the same task, each exploring different model architectures, feature engineering techniques, or parameter configurations. \projectName could then select the one that achieves the best performance on the evaluation dataset across multiple metrics, which could be optimized to execute in parallel but comes at an increased xApp generation cost. 

\subsection{Automated xApp Synthesis}
\label{subsec:xapp_syn}
After validating the core algorithms and functional components with either public or local datasets, the final step is to integrate them into a deployable xApp. 
\update{Unlike general-purpose program synthesis, O-RAN xApps must satisfy strict requirements on functional correctness, interface correctness, timing feasibility, and runtime stability.}
Conventional LLM-based code generation often stops once the code is syntactically correct or shows good offline accuracy, but \projectName treats \textit{validation} as a central objective and evaluates the correctness, robustness, timing behavior, and near-RT deployability of the generated control logic. Nevertheless, the complex nature of O-RAN platforms and dynamic working environments present significant practical challenges. 

\noindent\textbf{Technical challenge: deployable interface integration.} 
\update{One key challenge is \textit{interface matching}, which requires xApps to use the applicable E2 interface procedures, E2AP message structures, and E2 service-model definitions (\eg, E2SM-KPM for KPM reporting and E2SM-RC for RAN control)~\cite{ORAN_E2SM_COMMON}, and to encode control messages in corresponding formats.}
Any mismatch between the expected and available metrics--such as expecting per-UE throughput when only cell-level aggregates are reported--can result in functional failures. Moreover, \textit{policy enforcement} presents another layer of complexity. xApps must respect dynamic control rules defined by the RIC, such as slicing constraints, control priorities, and resource allocation limits. Even if an anomaly is correctly detected, xApps must verify whether intended actions (\eg, resource reallocation) are permissible under current policies. 
\update{\textit{Compatibility validation} is also essential to ensure that generated xApps satisfy runtime constraints such as configuration schemas, protocol versions, and execution-time bounds.}
For example, if the model inference of one xApp exceeds the 1-second near-RT latency requirement, it becomes inapplicable regardless of its correct logic or high accuracy.
\textit{These challenges collectively highlight the need for robust validation, adaptive interfacing, and policy-awareness mechanisms in the final deployment phase.}

\noindent\textbf{Design choice: platform-specific synthesis and validation.}
To tackle these challenges, \projectName constructs a platform-specific xApp template for the target RIC implementation, with OSC as the primary target. It does not assume that one generated xApp can run unchanged across different RIC implementations. The OSC-specific xApp follows OSC runtime, messaging, subscription, and lifecycle conventions. For other RIC implementations, such as FlexRIC~\cite{schmidt2021flexric}, \projectName retrieves the corresponding SDK documents, service-model definitions, and sample xApps as workflow and interface references, and regenerates the platform-specific interface layer. The template contains the components required for runtime initialization, configuration parsing, E2 message handling, algorithm execution, optional A1 policy enforcement, and output delivery. The validated core algorithm is inserted into this template to process incoming KPMs or other measurements and produce monitoring or control outputs. When the selected RIC implementation supports the required control capability, the output is encoded and sent to the RAN through the corresponding platform interface. The final xApp is produced through three stages: interface matching, algorithm integration, and xApp validation.

\begin{cmt}{Interface Matching}{}
\textbf{*User Problem*}\\
\textless user input\textgreater\{...\}\textless/user input\textgreater\\
\textbf{*Target*}\\
To ensure the xApp communicates correctly with O-RAN components, please generate complete and specification-aligned interface functions according to the scaffold structure, retrieved domain knowledge, and user requirements.\\
\textbf{*Rules*}\\
- Refer to the knowledge base.\\
- The generated functions should follow the data formats and message structures defined in O-RAN specifications. \\
- All code must be complete and self-contained.
\end{cmt}

\textbf{Interface Matching.} \update{This stage maps the input and output requirements of the generated algorithm to the measurements and control interfaces available in the target RIC/RAN deployment. Based on the selected or default RIC platform, \projectName retrieves the relevant interface specifications and deployment information from the knowledge base. It then checks whether the measurements, reporting scope, service models, and control interfaces required by the generated algorithm are supported by the selected platform. For each algorithm input, \projectName identifies the corresponding KPM, reporting granularity, and E2 service model.}  
We use a few-shot in-context learning Prompt~5, which provides several representative interface-generation examples and instructs the LLM to generate interface-related code according to the retrieved knowledge and user requirements. \projectName generates a platform-specific interface layer for the selected RIC platform, rather than producing one platform-independent xApp for all RIC implementations. When the available information is incomplete or ambiguous, \projectName returns to requirement refinement and requests only the necessary clarification from the operator. If a required capability is unavailable, \projectName uses an operator-approved alternative, when available, or reports that the requested functionality cannot be supported by the selected platform. These examples serve only as workflow and interface references for understanding registration, subscription, message flow, KPM reporting, and control-output delivery, while \projectName regenerates the xApp implementation for each requirement rather than reusing or filling in an existing sample application.

\noindent\textbf{Algorithm Integration.}
The algorithms and functional modules generated by \projectName take the public or local datasets as input (\S~\ref{sec:function_design}), rather than the real-world data streams reported via O-RAN interfaces. Thus, we need to modify and adapt the algorithms before integrating them into an xApp. Specifically, we design Prompt 6 that instructs \projectName to replace the offline data loading and evaluation code with interface-driven input and output handling logic. This includes adapting input pipelines to consume KPMs from E2 messages, restructuring the output format to align with xApp control actions, and embedding the algorithm logic into a real-time loop.

\begin{cmt}{Algorithm Integration}{}
\textbf{*User Problem*}\\
\textless user input\textgreater\{...\}\textless/user input\textgreater\\
\textbf{*Target*}\\
To transform the algorithm into a deployable xApp component, please adapt the code to read real-time E2 KPM inputs, process them with the existing algorithm logic, and output actionable results to the RIC system using standardized control message formats.\\
\textbf{*Rules*}\\
- Refer to the knowledge base.\\
- Follow the previous code template \{...\}.\\
- Avoid any blocking operations. \\
- All code must be complete with no placeholders.
\end{cmt}

\noindent\textbf{xApp Validation.}
To enhance the reliability of the integrated xApp, we further design an auxiliary function that supports essential tasks such as data parsing, logging, and configuration management.
In addition to syntax checking, \projectName uses SonarQube~\cite{sonar}, a static code analysis tool, to inspect the generated source code before deployment. It detects code-level issues such as bugs, code smells, maintainability problems, security warnings, and risky API or data-type usage, and feeds the reported issues back to the repair loop.
This process can effectively improve the reliability and maintainability of the generated xApp. However, it is important to note that the existing code analysis methods are limited to verifying code-level semantics and structural correctness. They do not assess the functional logic of the xApp. Specifically, these existing methods do not validate whether each interface is correctly implemented or real-time communication and control actions are performed as expected. 

To address this limitation, we perform system-level validation in the experimental evaluation (\S\ref{sec:realtesting}), where generated xApps are deployed and executed on a real-world O-RAN testbed. This runtime validation checks whether the generated xApp can be successfully deployed and executed, whether the requested KPMs can be received and decoded at the expected scope, whether required control interfaces, if any, are available, and whether the execution latency satisfies the near-RT requirement. In addition, \projectName performs runtime behavioral checks using known-input validation cases derived from collected KPM traces or task-specific test scenarios. The generated xApp is executed on these known inputs, and its outputs are checked against task-specific expectations, such as correctly detecting ground-truth anomalies while limiting false alarms and missed detections. If the validation exposes an unsupported measurement, an unexpected reporting granularity, an interface mismatch, or inconsistent runtime behavior, the corresponding logs are fed back to the Interface Matching or Algorithm Design stage for correction. If the mismatch cannot be resolved without changing the intended functionality, \projectName requests operator confirmation or reports that the functionality is unsupported in the selected deployment.
\section{Automated Deployment and Execution}

We now present the xApp deployment and execution process in our main OSC-based implementation. The xApp execution module is tightly coupled with O-RAN, ensuring that the entire workflow from xApp generation to deployment can be validated end-to-end within the operational O-RAN environment.

\noindent\textbf{Inter-unit Communication.} 
The generated xApps are deployed on the near-RT RIC. 
These xApps communicate internally via RIC Message Router (RMR)~\cite{santos2025managing} to handle inter-component messaging within the RIC software stack. The xApps communicate with other units in O-RAN via the standardized E2 interface and E2 Application Protocol (E2AP)~\cite{ORAN_E2AP}.
In particular, the xApps receive real-time telemetry from base stations (\eg, gNB) and transmit control messages to dynamically influence network behavior. 

\noindent\textbf{Automated xApp Deployment.} The generated xApps are packaged as Docker containers following the OSC xApp deployment workflow and are deployed onto the RIC host. Deployment scripts include all necessary environment configurations to enable integration with platform services such as logging, status monitoring, and configuration management. Containerized xApps are then launched by the RIC orchestrator and automatically establish initial connections to the RMR, subscription manager, and internal databases.

\noindent\textbf{Dynamic Registration Procedure.} Once deployed, the xApps register their presence and functional capabilities with the E2 Service Manager of the RIC.
This involves subscribing to specific metrics (\eg, RLC buffer occupancy, PRB utilization, handover statistics) emitted by the base stations, depending on the functionalities of the generated xApps. For instance, in the case of an anomaly detection xApp, \projectName ensures that only the relevant data streams are subscribed, minimizing telemetry transmission overhead.

\noindent\textbf{Real-time Execution and Decision Logic.} When the xApps operate in a closed-loop control mode, telemetry data collected via the E2 interface are continuously streamed into the xApps' processing engine. The core algorithms generated by \projectName process these data streams in real time for various tasks such as anomaly detection, load balancing, and interference mitigation. Based on real-time processing results, control actions (\eg, RAN parameter reconfiguration, traffic steering) are generated and sent back to the base stations via E2 to adapt network configurations.

\begin{figure}[t]
    \centering
    \setlength{\abovecaptionskip}{0pt}
    \includegraphics[width=0.85\linewidth]{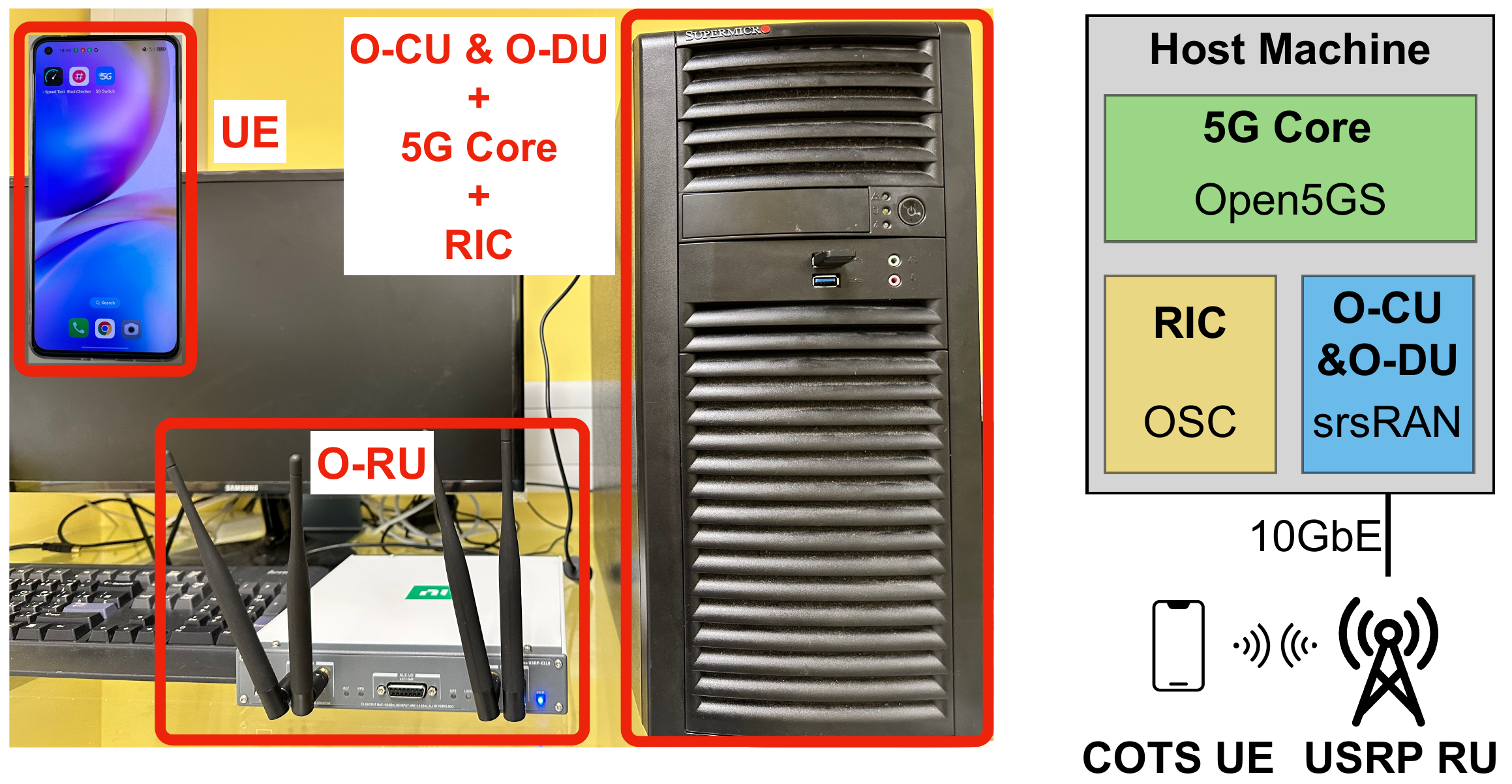}
    \caption{Testbedding with SDR and COTS UE.}
    \label{fig:experiment_setup}
    \vspace{-10pt}
\end{figure}


\noindent\update{\textbf{Standard O-RAN Interface Integration.}}
\update{The generated xApps use standard O-RAN interfaces according to the relevant specifications and the capabilities supported by the selected RIC/RAN platform. To this end, \projectName automatically generates task-specific E2 Service Model (E2SM) handlers, such as E2SM-KPM for KPM reporting and E2SM-RC for RAN control messaging. The generated interface logic follows the corresponding message formats, telemetry structures, and control-interface definitions. We further validate the resulting xApps through deployment and runtime communication checks on the evaluated platforms. These checks demonstrate interface and runtime compatibility under the tested configurations, but do not constitute formal O-RAN conformance testing or certification. This also does not assume source-level portability across different RIC implementations.}
\section{Implementation and Evaluation}
\label{sec:evaluation}

In this section, we introduce the implementation and evaluation of \projectName, aiming to answer the following key research questions: 
(1) Can \projectName generate xApps that satisfy the specified task objectives and achieve task-level performance comparable to \update{manually developed} baselines?
(2) Can \projectName-generated xApps be automatically deployed and executed on real-world O-RAN platforms with near-RT control loop constraints?
(3) How does each novel technical module contribute to the overall performance of  \projectName?


\subsection{Experiment Setup}
\label{sec:setup}


\textbf{Hardware.}
We implement \projectName on the software-defined srsRAN~\cite{srsRAN} stack.
As shown in~\figurename~\ref{fig:experiment_setup}, we use a Ubuntu 22.04.1 LTS workstation with an Intel Xeon(R) Core E5-2620 v4 CPU, 32 GB RAM, and USRP X310 units to serve as the gNodeB. 
We run Open5GS~\cite{Open5GS} on the workstation as the core network. User equipments (UEs) include OnePlus 8T and Xiaomi 13 Pro smartphones with programmable SIM/USIM cards (sysmoISIM-SJA2 SIM cards~\cite{sysmoISIM}). 
For control-plane support, we use the OSC near-RT RIC as the main xApp execution platform to host generated xApps and manage near-RT RAN control. In this testbed, srsRAN provides the RAN-side gNB/E2-node functionality, while the generated xApps run on the OSC near-RT RIC and interact with the srsRAN-based RAN stack through KPM subscription, KPM reception/parsing, and monitoring or control execution.

\noindent\textbf{Software.}
We use GPT-4o~\cite{achiam2023gpt,hurst2024gpt} as the default LLM and LangChain~\cite{langchain} as the orchestration framework for prompt management, module integration, and tool wrapping. We implement all backend logic in Python 3.10 and utilize libraries such as OpenAI SDK, FAISS~\cite{douze2025faiss}, and FastAPI~\cite{fastapi}. Specifically, LangChain coordinates prompt templates, LLM calls, retrieval utilities, and validation tools across different generation modules; FAISS provides embedding-based similarity search over the local O-RAN knowledge base; and FastAPI exposes lightweight REST APIs for user interaction, module invocation, and backend service integration. SonarQube~\cite{sonar} is used for static code analysis.

\vspace{-2pt}
\subsection{Open-Source Dataset Performance}
\label{sec:performance}
\vspace{-2pt}
We compare \projectName with three representative baselines to evaluate whether \projectName-developed xApps can satisfy different user requirements across input types and xApp tasks. The three baselines are organized around two evaluation settings.

\noindent\textbf{Anomaly detection under different input pathways.}
SpotLight~\cite{sun2024spotlight} and MobiWatch~\cite{wen20246g} share the same high-level objective of anomaly detection, but rely on different input pathways. SpotLight uses fine-grained RAN KPM telemetry, where the generated xApp selects representative measurements such as PRB utilization, throughput, latency, packet loss, and RSRP from the available E2SM-KPM telemetry according to the required reporting scope. In contrast, MobiWatch uses link-layer and session-layer signaling traces rather than E2SM-KPM counters. This setting evaluates whether \projectName can generate functionally similar xApps under different developer requirements on input data.

\noindent\textbf{Generalization to radio-layer interference classification.}
IC~\cite{chiejina2024system} represents a different radio-layer task, i.e., interference classification, and is used to examine whether \projectName can generalize beyond anomaly detection to another xApp function and data modality. For the KPM-based IC setting, the generated xApp selects radio-link measurements such as SINR, BLER, MCS, throughput, and PRB utilization from the available E2SM-KPM telemetry according to the required reporting scope.

For fair comparisons, we use the same dataset (\ie, the one that each baseline uses) for evaluation. For \projectName-generated xApps, the datasets are used in the algorithm validation loop to evaluate and refine the generated algorithms, rather than to fine-tune the LLM. To ensure a fair comparison, we use the same fixed 4:6 train/test split for \projectName and the baseline algorithms, and report all final results on the testing set.

\subsubsection{Evaluation Metrics}

We adopt two sets of metrics:

\noindent\textbf{\projectName-generated xApp Evaluation.} (1) \textit{Precision}, \textit{Recall Rate} and \textit{F1 Score} are the primary indicators of model correctness. (2) \textit{VRAM Usage} evaluates the runtime efficiency of the generated AI/ML algorithms.



\begin{table}[t]
\centering
\setlength{\abovecaptionskip}{4pt}
\caption{Accuracy comparison with SpotLight \cite{sun2024spotlight}}
\resizebox{0.45\textwidth}{!}{\begin{tabular}{c c | c c c c c}
\specialrule{0.2em}{2.5pt}{2.5pt}
\textbf{Method} & Metric &\textbf{MAC} & \textbf{NETWORK} & \textbf{PDCP}& \textbf{RADIO} & \textbf{MIXED}\\
\specialrule{0.2em}{2.5pt}{2.5pt}
\multirow{2}{*}{\textbf{\projectName}}&Precision& 97.3\% & 98.9\% & 92.1\% & 78.8\% & 97.6\% \\
& Recall & 97.5\% & 98.9\% & 91.8\% & 81.5\% & 97.6\%\\
\specialrule{0.2em}{2.5pt}{2.5pt}
\multirow{2}{*}{SpotLight}&Precision& 93.6\% & 94\% & 100\% & 95\% & 95.5\% \\
& Recall & 100\% & 92\% & 93\% & 93\% & 94.5\% \\
\specialrule{0.1em}{2.5pt}{2.5pt}
\multirow{2}{*}{Z Score}&Precision& 74.4\% & 75\% & 69.2\% & 58.8\% & 65.4\%\\
& Recall & 82.4\% & 83.6\% & 69.4\% & 59.7\% & 79.2\% \\
\specialrule{0.1em}{2.5pt}{2.5pt}
\multirow{2}{*}{LSTM}&Precision& 74.5\% & 65.2\% & 73.5\% & 92.05\% & 69.5\% \\
& Recall & 6.3\% & 37.6\% & 7.1\% & 54.2\% & 10.5\% \\
\specialrule{0.2em}{2.5pt}{2.5pt}
\end{tabular}}
\vspace{-10pt}
\label{tab:SpotLight_Acc}
\end{table}

\noindent\textbf{\projectName Evaluation.}
(3) \textit{Synthesis Time} quantifies the total duration from the moment a user requirement is input to the completion of xApp generation;
(4) \textit{Number of Bugs} assesses the code quality and the robustness of \textit{xAppGen}’s xApp generation. Each syntactic or semantic error caught during compilation or testing is counted as a bug;
(5) \textit{Iteration-to-Success Count} measures the correction cycles required to synthesize a fully functional program;
(6) \textit{One-Shot Success Rate} measures the proportion of xApp generation tasks that can be completed successfully in the first attempt, from the operator's high-level intent to a deployed and executable xApp, without additional repair or regeneration iterations.

\subsubsection{Baseline 1 - SpotLight for Anomaly Detection}
\label{sec:baseline1}

SpotLight collects a comprehensive 5G O-RAN dataset, containing over 100 million KPM datapoints sampled at 100ms resolution. The dataset covers 600+ metrics across MAC, RLC, PDCP, Radio, and platform components, and captures both synthetic and real-world anomalies such as CPU contention and fronthaul congestion. SpotLight adopts a two-stage distribution learning method (\ie, JVGAN+MRPI) to detect the anomalous behaviors of RAN KPMs. For fair comparison, we leverage \projectName to synthesize an anomaly detection xApp based on the user requirements. We also report the performance of a typical Z-score detection algorithm and an LSTM-based autoencoder for comparison.

\noindent\textbf{Results.} \tableautorefname{}~\ref{tab:SpotLight_Acc} lists the precision and recall rate on five KPM subsets: MAC, Network, PDCP, Radio, and Mixed, corresponding to five different O-RAN anomalies. We observe that \projectName-developed xApps achieve higher precision and recall rates across all subsets. Surprisingly, it surpasses SpotLight in several subsets, especially when using Mixed KPMs that aggregate multiple cases. Further analysis of the generated code reveals that \projectName can automatically select and preprocess the most relevant KPMs, while discarding noisy or weakly correlated knowledge during prompt construction. \projectName also applies task-specific filtering strategies during data preparation, leading to more robust features and better generalization across anomaly types.

\begin{figure}[t]
\setlength{\abovecaptionskip}{0pt}
\subfigtopskip=-4pt
\subfigcapskip=-4pt
	\centering
	\subfigure[Quality vs. Iterations]{
		\includegraphics[width=0.45\linewidth]{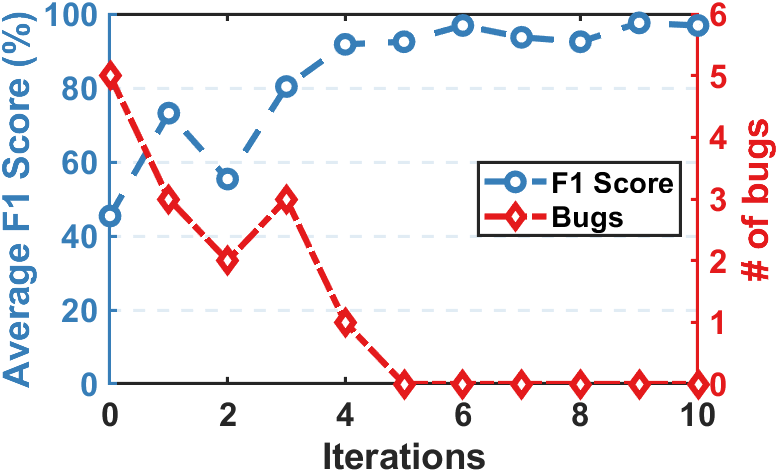}}
	\hfill
	\subfigure[Model diversity]{
		\includegraphics[width=0.45\linewidth]{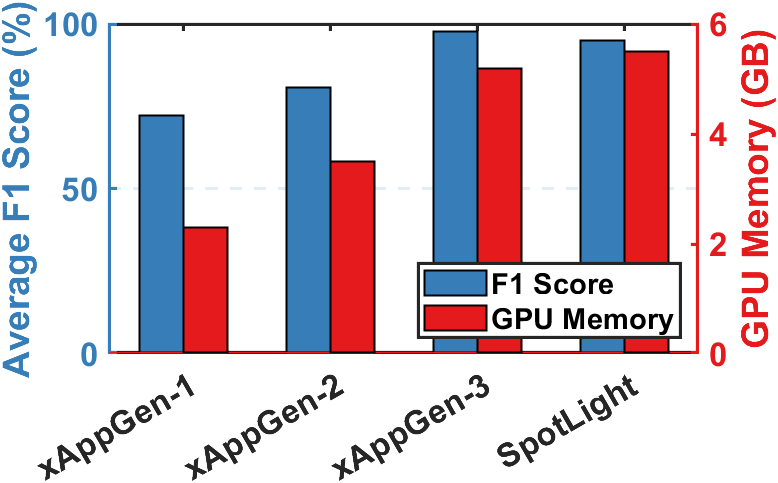}}

	\caption{Performance of \projectName-developed anomaly detection xApp on SpotLight dataset: (a) quality improvement across iterations; and (b) xApp variants.}
    \vspace{-10pt}
	\label{fig:SpotLight_Bug}
\end{figure}

To examine the model iteration process of \projectName, we analyze both accuracy and code quality across multiple design and validation cycles. As shown in \figurename~\ref{fig:SpotLight_Bug}(a), the generated xApps achieve higher accuracy and fewer bugs after iterative refinement. On average, the initial implementation contains approximately five bugs per trial across 25 independent runs, mainly due to incomplete logic, improper data preprocessing, and incorrect control flow. Notably, after two to three iterations, the number of bugs decreases significantly, while the accuracy also increases. This shows that the code improvement module can use prior errors and system feedback to improve both functional correctness and task performance.

Beyond accuracy, \projectName also supports interactive customization. As shown in \figurename~\ref{fig:SpotLight_Bug}(b), the three generated variants (\ie, \textit{\projectName-1} to \textit{\projectName-3}) differ in feature selection, model architecture, accuracy, and VRAM usage. This allows operators to choose an xApp variant according to available hardware resources and performance requirements.

\begin{figure}[t]
\setlength{\abovecaptionskip}{0pt}
\subfigtopskip=0pt
\subfigcapskip=0pt
	\centering
	\subfigure[Quality vs. Iteration]{
		\includegraphics[width=0.45\linewidth]{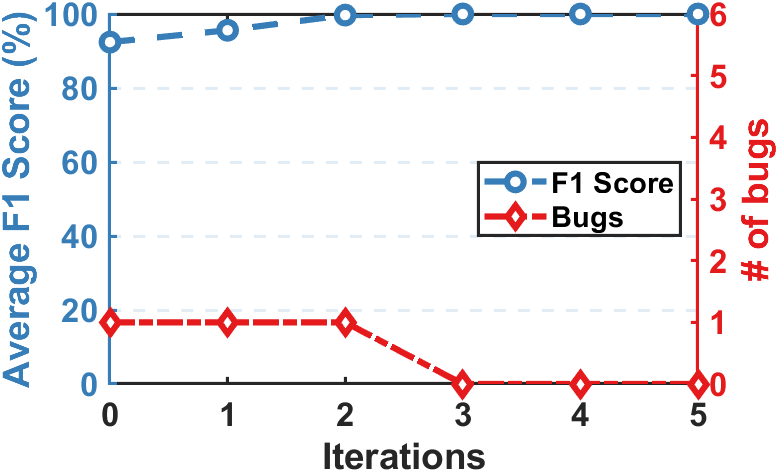}}
	\hfill
	\subfigure[Model diversity]{
		\includegraphics[width=0.45\linewidth]{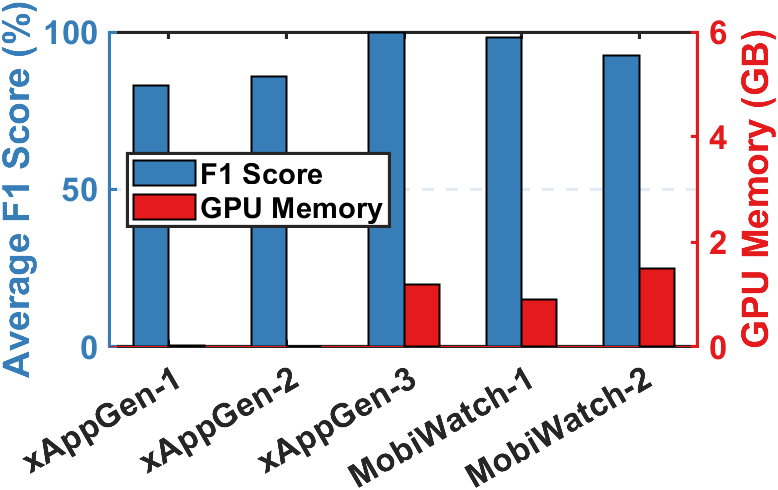}}

	\caption{Performance of \projectName-developed anomaly detection xApp on MobiWatch: (a) quality increment across iterations; and (b) diverse xApp variants.}
	\label{fig:MobiWatch_Bug}
        \vspace{-10pt}
\end{figure}

\subsubsection{Baseline 2 - MobiWatch for Anomaly Detection} MobiWatch monitors link-layer and session-layer (RRC and NAS) messages using MobiFlow \cite{wen20246g}, a telemetry pipeline that transforms packet traces into structured signaling features. The dataset includes both benign and adversarial traces. MobiWatch implements two versions of anomaly detection using Autoencoders \cite{rumelhart1985learning} and LSTM \cite{hochreiter1997long}, noted as MobiWatch-1 and MobiWatch-2, respectively. 

\noindent\textbf{Results.} \tableautorefname{}~\ref{tab:MobiWatch_Acc} reports the average accuracy, precision, and recall of both MobiWatch and \projectName-developed xApps. Since the input feature space of MobiWatch is more structured and lower-dimensional than the heterogeneous KPM telemetry used in Baseline~1, the accuracy of \projectName significantly outperforms MobiWatch on the same dataset. \figurename{}~\ref{fig:MobiWatch_Bug}(a) illustrates the model iteration process. We observe that for simpler tasks, \projectName can iterate more quickly to a satisfactory model with $<$3 iterations. In addition, \projectName provides various xApp options as shown in \figurename{}~\ref{fig:MobiWatch_Bug}(b). Users can choose simple models (\ie, \textit{\projectName-2} without GPU usage) to achieve comparable performance.

\subsubsection{Baseline 3 - IC for Interference Classification} 
This dataset was collected in \cite{chiejina2024system} using a real srsRAN-based O-RAN testbed. It includes two modalities: 10,000 spectrograms (128$\times$128 grayscale) recorded over the air from USRP-based receivers, and over 25,000 uplink KPM traces including SINR, BLER, MCS, and throughput metrics. Note that half of the samples correspond to clean transmission, while the other half represent continuous-wave jamming scenarios.
\begin{figure*}[t]
\setlength{\abovecaptionskip}{0pt}
\subfigtopskip=0pt
\subfigcapskip=0pt
	\centering
	\subfigure[\projectName - Baseline 1]{%
		\label{fig:AutORAN_SAD_llm}
		\includegraphics[width=0.245\linewidth]{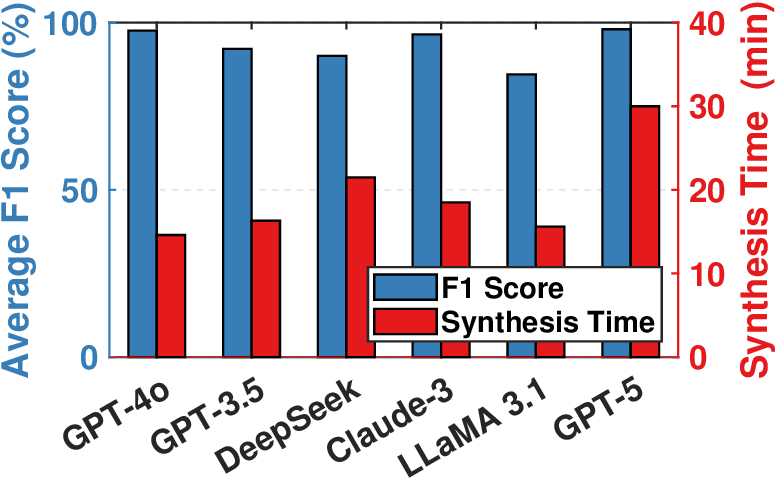}}%
	\hfill
    \subfigure[\projectName - Baseline 2]{%
		\label{fig:AutORAN_MAD_llm}
		\includegraphics[width=0.245\linewidth]{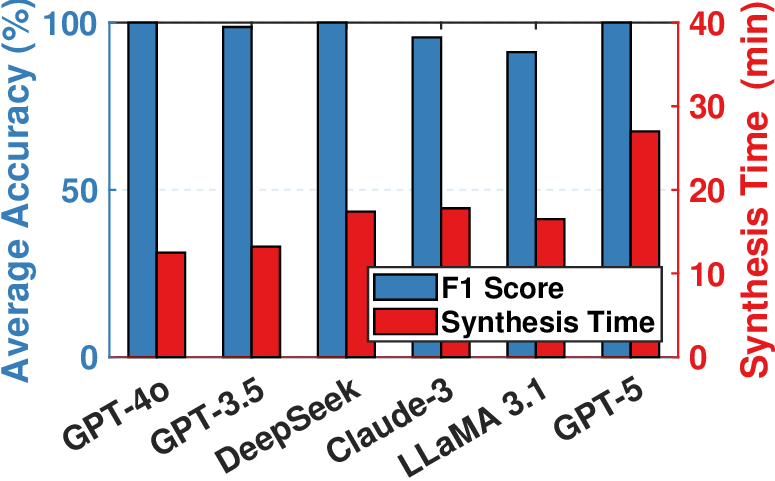}}%
    \hfill
	\subfigure[\projectName - Baseline 3]{%
		\label{fig:AutORAN_IC_llm}
		\includegraphics[width=0.245\linewidth]{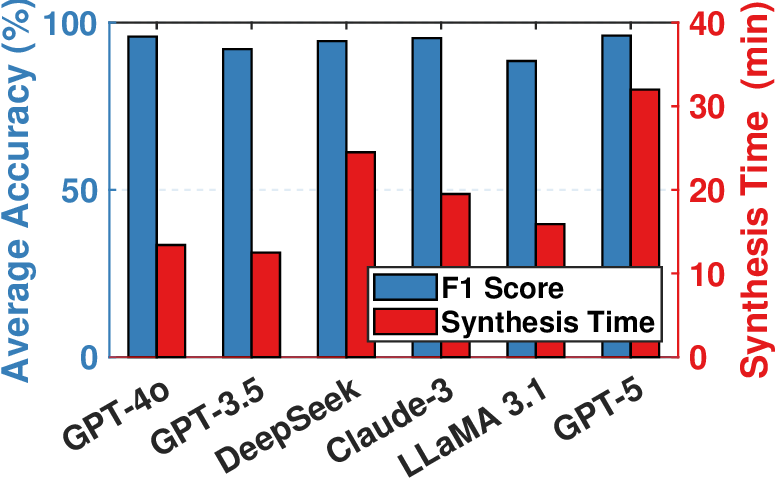}}%
    \hfill
	\subfigure[Programming language]{%
		\label{fig:Language}
		\includegraphics[width=0.23\linewidth]{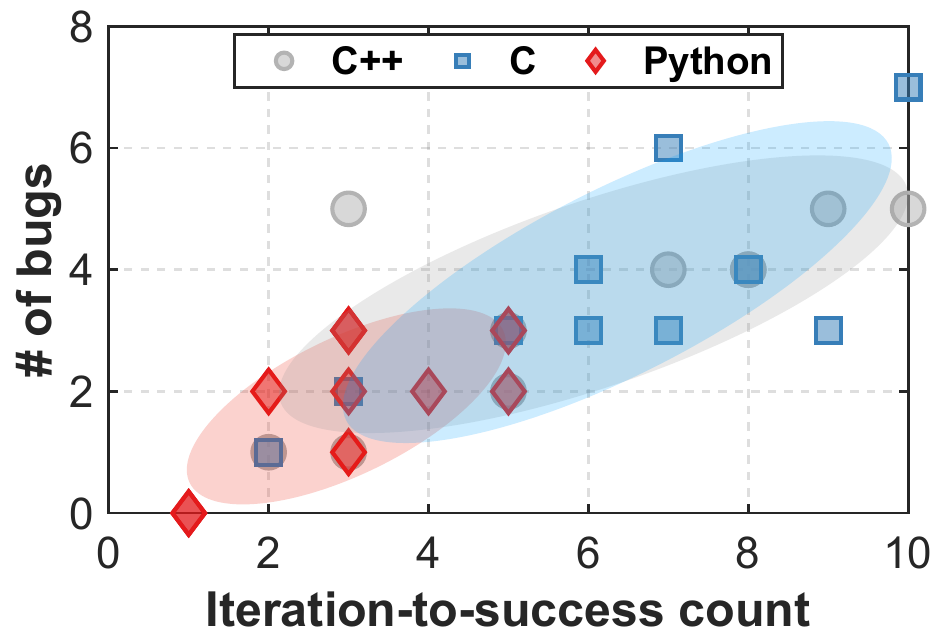}}%
	\caption{Impact of different LLM and programming languages on \projectName-generated xApps.}
	\label{fig:different_llm}
         \vspace{-10pt}
\end{figure*}

\begin{table}[t]
\centering
\small
\caption{Accuracy comparison with IC \cite{chiejina2024system}}
\begin{tabular}{c c | c}
\specialrule{0.2em}{2.5pt}{2.5pt}
\textbf{Method} & Type of xApps (Dataset) &\textbf{Accuracy} \\
\specialrule{0.2em}{2.5pt}{2.5pt}
\multirow{2}{*}{\textbf{\projectName}}&InterClass-Spec& 92.9\% \\
& InterClass-KPM & 98.8\%  \\
\specialrule{0.1em}{2.5pt}{2.5pt}
\multirow{2}{*}{IC~\cite{chiejina2024system}}&InterClass-Spec& 98\% \\
& InterClass-KPM & 97.9\% \\
\specialrule{0.2em}{2.5pt}{2.5pt}
\end{tabular}
\vspace{-10pt}
\label{tab:precision_recall_xapp3}
\end{table}

\noindent\textbf{Results.}
We generated two versions of \projectName-IC xApps using either the spectrogram data (InterClass-Spec) or KPM subsets (InterClass-KPM). As shown in \tableautorefname~\ref{tab:precision_recall_xapp3}, \projectName achieves detection accuracies of 92.9\% and 98.8\% on the two respective subsets, matching the performance of the CNN and DNN models proposed in the baseline paper \cite{chiejina2024system}. However, we observed significant performance variance across different generated versions: some models achieve less than 80\% accuracy, while others exceed 95\%. The underlying reasons are twofold. Unlike time-series KPM data, spectrograms are high-dimensional images, which require more complex algorithms for effective and robust data preprocessing. In addition, inappropriate selections of AI/ML model architectures and hyperparameters from the vast configuration space can lead to unstable performance\cite{li2020rethinking}. To tackle this issue, we may further integrate architecture-level reasoning and guide search strategies into the generation process. 
Overall, the experiment results demonstrate that \projectName is capable of generating xApp functions for radio-layer tasks by adopting more complex image-based AI/ML models.

\subsection{Key Impact Factors}
\label{sec:key_impact_factors}


Beyond task-level performance, we further examine key factors that affect the quality, efficiency, and deployability of \projectName-generated xApps, including LLM choice, programming language, generation cost, and target RIC implementation. These analyses complement the task-level accuracy results and provide process-level evidence on the practical use of \projectName for automated xApp development.

\subsubsection{Impact of LLM Choice}
We test six LLMs for comparison: GPT-4o~\cite{achiam2023gpt,hurst2024gpt}, GPT-3.5~\cite{brown2020language}, DeepSeek-R1~\cite{liu2024deepseek}, Claude 3~\cite{Claude}, LLaMA 3.1 (70B)~\cite{touvron2023llama} and GPT-5~\cite{gpt5}. Each model was used to independently generate xApps, which were subsequently evaluated in terms of accuracy/F1 score and total synthesis time. As shown in \figurename~\ref{fig:AutORAN_SAD_llm}, the xApp generated via GPT-4o consistently achieves strong performance across all datasets, while GPT-5 delivers nearly identical accuracy but with longer synthesis time. This superior performance stems from three key factors: (1) stronger chain-of-thought reasoning, leading to better decomposition of high-level xApp logic; (2) more semantically aligned code with fewer hallucinations or misinterpretations of O-RAN domain terms; and (3) more modular and syntactically complete outputs, reducing post-generation correction.

Moreover, GPT-4o-based \projectName requires less than 20 minutes to generate most xApps, while GPT-5 achieves comparable accuracy with longer wall-clock synthesis time. Other models show greater variability, often producing verbose code, relying on overly generic patterns, or omitting low-level implementation details. As shown in \figurename~\ref{fig:different_llm}, LLM performance also varies across tasks: structured KPM-based anomaly detection is generally easier to synthesize, whereas tasks involving fine-grained telemetry or high-dimensional spectrogram inputs are more challenging. For Baseline~2, GPT-3.5 and DeepSeek-based variants often misinterpret the input schema, leading to accuracy drops of 10--15\% compared with GPT-4o and GPT-5. For Baseline~3, models that fail to generate valid spectrogram preprocessing stages perform substantially worse. These results show that xApp synthesis quality depends on both the input modality and the LLM's ability to align generated code with O-RAN-specific structures and constraints.

\subsubsection{Impact of Programming Language}
We explicitly prompted \projectName to synthesize functionally equivalent xApps in Python, C, and C++. Each version was tested across 10 independent trials using the same requirements and input dataset. We then recorded the cumulative bug count before the first successful run and the iteration-to-success count to assess the code reliability and structural difficulty. As shown in \figurename~\ref{fig:Language}, Python xApps exhibit higher efficiency and concise structures, averaging only 1.8 bugs per version and requiring only 3 iterations to achieve a fully functional implementation. This advantage can be attributed to Python's high-level syntax, which resembles natural language compared to C and C++. Its concise, human-readable structure enables LLMs to more effectively translate high-level user requirements into functional code \cite{austin2021program}. In contrast, C and C++ based xApps exhibit more bugs -- often due to errors in memory management, pointer arithmetic, and data type handling -- and require more code refinement iterations to converge.

\subsubsection{Generation Cost}
We also estimate the generation cost of \projectName under the GPT-4o pricing~\cite{Openaiprice} used in our experiments. Based on our multi-stage prompt workflow and the generation/repair process, generating one usable xApp consumes roughly 1.0M--2.5M tokens in total, corresponding to approximately \$7--\$15 per usable xApp. This cost varies with xApp complexity, since more complex xApps typically involve larger context inputs, longer generated code, more validation logs, and additional repair iterations.

\subsubsection{Multi-target Generation and RIC Communication Validation}
\label{sec:multi_target}
To clarify the portability boundary and runtime communication capability of \projectName, we conduct a multi-target generation case study based on the same high-level requirement, which is to generate a KPM-based anomaly detection xApp. We instantiate this requirement for two target RIC implementations, OSC and FlexRIC\footnote{\update{The generated OSC-specific and FlexRIC-specific xApps are publicly available at \url{https://anonymous.4open.science/r/xAppGen/}.}}, and evaluate whether the generated xApps can preserve the same algorithmic functionality while adapting to different RIC runtimes and interfaces. \update{Both implementations use E2SM-KPM (RAN Function ID 2), with Report Styles 1 and 4 and a 1000-ms reporting period. The subscribed KPMs include \texttt{RRU.PrbUsedDl} and \texttt{RRU.PrbAvailDl}, which are used together to derive PRB utilization rather than directly using \texttt{RRU.PrbTotDl}, as well as \texttt{DRB.UEThpDl}, \texttt{DRB.RlcSduDelayDl}, and \texttt{DRB.RlcPacketDropRateDl}.}

Table~\ref{tab:multi_target} reports both algorithm-level performance and platform-level communication validation results. We do not claim formal O-RAN conformance certification. 
\update{Instead, we perform interface and runtime validation checks to verify whether the generated xApps can complete the runtime workflow required by the target RIC implementation, including deployment, RIC-side registration, subscription setup, KPM reception, KPM parsing, alarm/control-output generation, end-to-end execution, and runtime trace inspection. Passing these checks demonstrates interface and runtime compatibility with the tested RIC implementations and configurations; it does not constitute formal O-RAN conformance testing or certification.}
Since the two xApps implement the same anomaly detection objective, they achieve the same F1 score, while their execution-time ranges and interface layers differ due to platform-specific runtime and messaging logic. Both generated xApps pass the corresponding communication and runtime validation checks. These results show that \projectName maps the same high-level requirement to different platform-specific xApp implementations by regenerating the platform-dependent interface layer, rather than relying on source-level portability of a single xApp.

\begin{table}[t]
\centering
\caption{Multi-target generation and RIC communication validation results from the same high-level requirement.}
\label{tab:multi_target}
\resizebox{\columnwidth}{!}{
\begin{tabular}{lcc}
\specialrule{0.2em}{2.5pt}{2.5pt}
\textbf{Metric} & \textbf{OSC-specific xApp} & \textbf{FlexRIC-specific xApp} \\
\specialrule{0.2em}{2.5pt}{2.5pt}
Performance (F1 score) & 97.6\% & 97.6\% \\
Execution time & 20--110 ms & 18--95 ms \\
Deployment success & $\checkmark$ & $\checkmark$ \\
RIC-side registration & $\checkmark$ & $\checkmark$ \\
Subscription setup & $\checkmark$ & $\checkmark$ \\
KPM reception & $\checkmark$ & $\checkmark$ \\
KPM parsing correctness & $\checkmark$ & $\checkmark$ \\
Alarm/control-output generation & $\checkmark$ & $\checkmark$ \\
End-to-end execution & $\checkmark$ & $\checkmark$ \\
Runtime trace inspection & $\checkmark$ & $\checkmark$ \\
\specialrule{0.2em}{2.5pt}{2.5pt}
\end{tabular}
}
\vspace{-8pt}
\end{table}

\subsection{Private \& Domain-Specific Performance}
\begin{table}[t]
\centering
\scriptsize
\setlength{\abovecaptionskip}{4pt}
\caption{Accuracy comparison with MobiWatch \cite{wen20246g}}
\begin{tabular}{c c | c c c}
\specialrule{0.2em}{2.5pt}{2.5pt}
\textbf{Method} & Dataset &\textbf{Accuracy} & \textbf{Precision} & \textbf{Recall}\\
\specialrule{0.2em}{2.5pt}{2.5pt}
\multirow{2}{*}{\textbf{\projectName}}&Benign& 100\% & 100\% & N/A \\
& Attack & 100\% & 100\% & 100\% \\
\specialrule{0.1em}{2.5pt}{2.5pt}
\multirow{2}{*}{MobiWatch-1}&Benign& 93.23\% & 93.23\% & N/A \\
& Attack & 100\% & 100\% & 100\% \\
\specialrule{0.1em}{2.5pt}{2.5pt}
\multirow{2}{*}{MobiWatch-2}&Benign& 91.15\% & 91.15\% & N/A \\
& Attack & 95\% & 88.68\% & 100\% \\
\specialrule{0.2em}{2.5pt}{2.5pt}
\end{tabular}
\vspace{-5pt}
\label{tab:MobiWatch_Acc}
\end{table}
To further assess the generalizability and practical utility of \projectName, we evaluate a slice scheduling xApp, which requires real-time radio resource allocation under multi-slice QoS constraints. Since public datasets for operator-specific slice scheduling are unavailable, we construct a synthetic dataset that emulates per-slice telemetry and QoS policies based on standardized O-RAN interfaces. This also avoids potential memorization effects from public samples and provides a controlled setting for evaluating the generalization capability.

\begin{figure}[t]
\setlength{\abovecaptionskip}{0pt}
\subfigcapskip=-4pt
	\centering
	\subfigure[Latency vs. throughput per slice]{
		\includegraphics[width=0.45\linewidth]{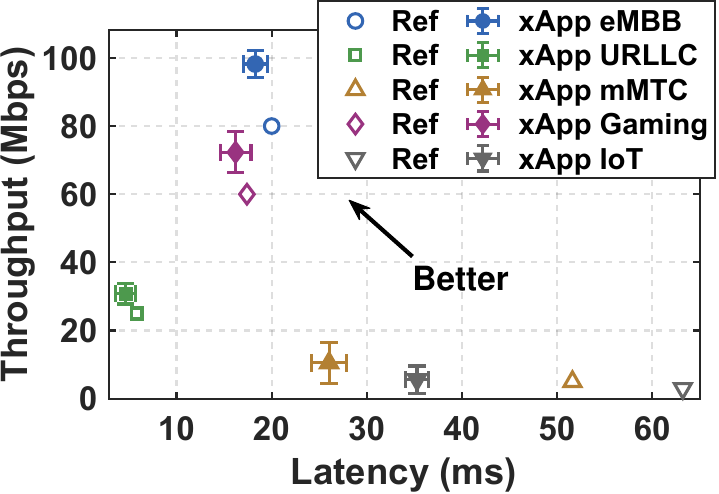} \label{fig:ss_latency}}
	\hfill
	\subfigure[QoS satisfaction over time]{
		\includegraphics[width=0.45\linewidth]{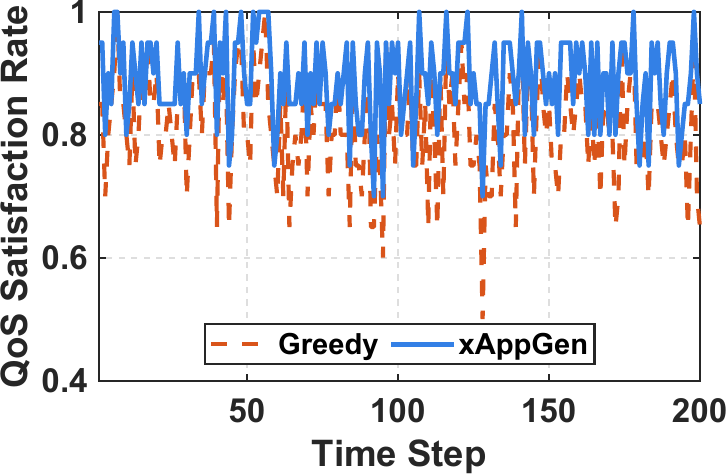}
        \label{fig:ss_compare}}
	\caption{Performance of \projectName-generated slice scheduling xApp.}
    \vspace{-5pt}
\end{figure}


We extend standard service types defined by 3GPP \cite{3GPP} to include five representative slice categories, where each class embodies distinct QoS objectives—eMBB emphasizes sustained high throughput, URLLC targets ultra-low latency, mMTC prioritizes energy-efficient massive connectivity, Gaming demands low latency with stable jitter, and IoT focuses on reliable transmission under low data-rate conditions. These slice definitions are instantiated across the O-RAN stack, influencing A1 policy configurations, guiding E2 node telemetry reporting, and serving as templates for SMO-driven service orchestration. To support evaluation, we construct a synthetic dataset that closely replicates slice-level telemetry patterns observed in operational O-RAN deployments.


The telemetry generation process follows the standardized E2SM-KPM service model and captures a set of representative per-slice metrics, including Physical Resource Block (PRB) utilization, end-to-end latency, throughput, packet loss, Reference Signal Received Power (RSRP), channel quality, modulation and coding information, and buffer status. Slice-specific QoS targets and relative priorities are derived from A1 policy profiles, enabling the dataset to encode both observed performance states and policy-driven intents.
For the generated slice scheduling xApp, \projectName produces radio-resource allocation decisions that identify the target slice and specify its PRB allocation ratio or resource quota. The xApp therefore belongs to the radio-resource allocation control type. During Interface Matching, these control outputs are mapped to the E2SM-RC control style and attributes supported by the selected RIC/RAN platform~\cite{ORAN_E2SM_RC}, according to the corresponding service-model definitions and platform capabilities. 
\update{E2SM-RC defines several Control Styles for different RAN functions, including Style~1 for Radio Bearer Control, Style~2 for Radio Resource Allocation Control, Style~3 for Connected Mode Mobility Control, Style~4 for Radio Access Control, and Style~5 for Dual Connectivity Control. The generated slice scheduling xApp uses Style~2 (Radio Resource Allocation Control), with Action ID~6, Control Header Format~1, and Control Message Format~1. Its control message identifies the target slice using PLMN and S-NSSAI (SST and SD) and specifies the minimum, maximum, and dedicated resource quotas.} 
Hence, \projectName does not hard-code a universal E2SM-RC style or attribute set across different platforms.
To enhance realism, we calibrate value ranges, inter-metric correlations, and noise models using empirical trends reported in prior measurement studies. In particular, SpotLight provides insights into PRB allocation and interference patterns, while MobiWatch reports user mobility and throughput dynamics. These studies guide the construction of slice-level telemetry patterns for the synthetic dataset.

We evaluate the \projectName-generated slice scheduling xApp by jointly examining its effect on per-slice throughput/latency trade-offs and on the temporal evolution of QoS satisfaction. As shown in \figurename{}~\ref{fig:ss_latency}, each marker pair represents the Requirement (Ref) and the \projectName-generated xApp for the same slice type. Across all five slice categories, the xApp consistently shifts the operating point towards the top-left region, indicating simultaneous throughput gains and latency reductions. URLLC slices, in particular, exhibit the largest latency improvement, confirming that \projectName can generate control logic that prioritizes delay-sensitive services while preserving throughput. \figurename{}~\ref{fig:ss_compare} further presents the QoS satisfaction rate over time under two scheduling strategies: a greedy heuristic baseline and \projectName. While the greedy approach achieves only moderate compliance and fluctuates heavily with traffic changes, the \projectName-generated xApp maintains consistently higher satisfaction rates by dynamically reallocating PRBs in response to real-time slice performance feedback, ensuring robust QoS across heterogeneous and time-varying loads.

\subsection{Real-World Execution Performance}
\label{sec:realtesting}
To distinguish dataset-level algorithm validation from full-stack runtime testing, we further conduct integration and runtime experiments on the srsRAN-based O-RAN testbed. The public and synthetic datasets used in previous evaluations mainly assess the core algorithmic performance of generated xApps, such as anomaly detection accuracy and slice scheduling effectiveness. In contrast, this subsection evaluates whether the generated xApps can be onboarded into the OSC near-RT RIC, subscribe to KPM streams exposed by the srsRAN-based RAN stack, receive and parse KPM messages, and execute their monitoring or control logic under near-RT runtime constraints. For each representative function (two anomaly detection applications and one slice scheduling application), we automatically generate three distinct algorithmic variants of the xApp, denoted as \projectName-1/2/3, and onboard them into the near-RT RIC platform via standard integration procedures. In addition to measuring execution latency, we also inspect whether the generated xApps produce expected runtime outputs under known input traces.


\begin{figure}[t]
\setlength{\abovecaptionskip}{0pt}
    \centering
    \includegraphics[width=0.9\linewidth]{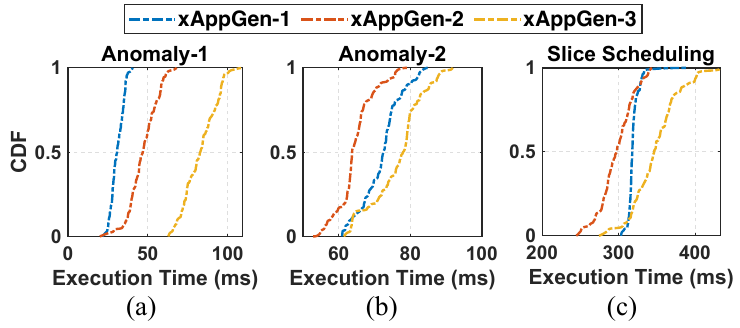}
    \caption{Execution time of \projectName-generated xApps on real-world testbed.}
    \label{fig:xappcdf}
    \vspace{-10pt}
\end{figure}

\textbf{Execution Time.} We measured the end-to-end execution time of the complete control loop, encompassing KPM stream reception, model inference, and control policy execution. As shown in \figurename~\ref{fig:xappcdf}(a,b), the two anomaly detection xApps exhibit execution times ranging from 20–110 ms and 55–90 ms, respectively. The slightly higher latency of the second baseline is attributed to its more intensive telemetry analysis. For slice scheduling (\figurename~\ref{fig:xappcdf}(c)), execution times span 250–420 ms, reflecting the added complexity of multi-slice QoS optimization.
These results demonstrate that \projectName-generated xApps are deployable within the sub-second latency constraints of near-RT RIC control loops, even when accommodating more sophisticated control tasks.

\begin{figure}[t]
    \centering
    \includegraphics[width=0.9\linewidth]{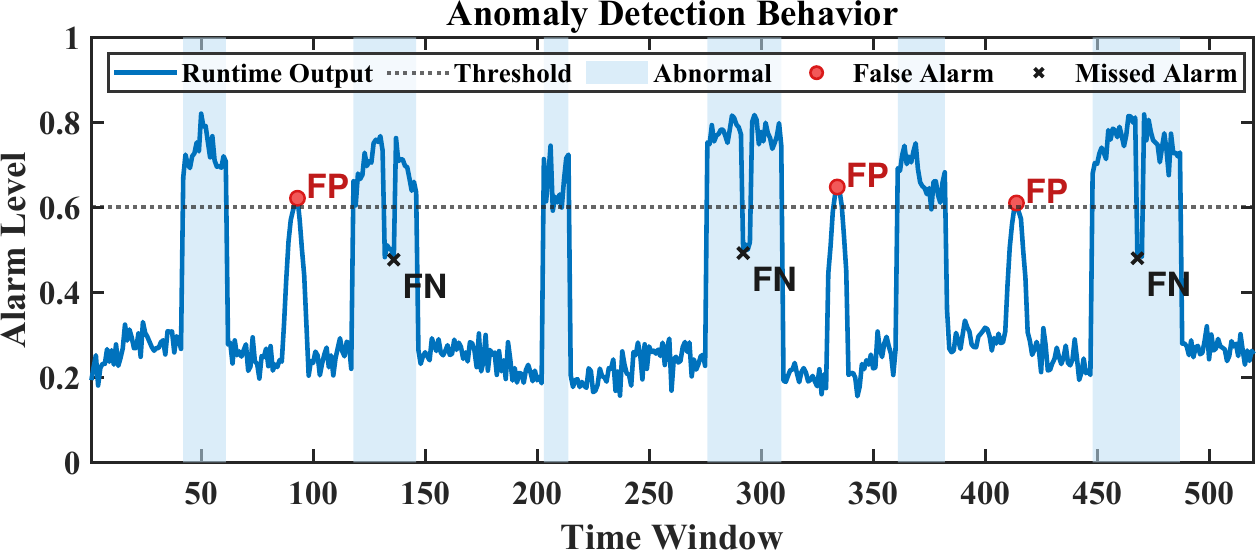}
    \caption{Runtime behavior of the generated anomaly detection xApp.}
    \label{fig:behavioral_test}
    \vspace{-5pt}
\end{figure}

\begin{figure}[t]
    \centering
    \includegraphics[width=0.9\linewidth]{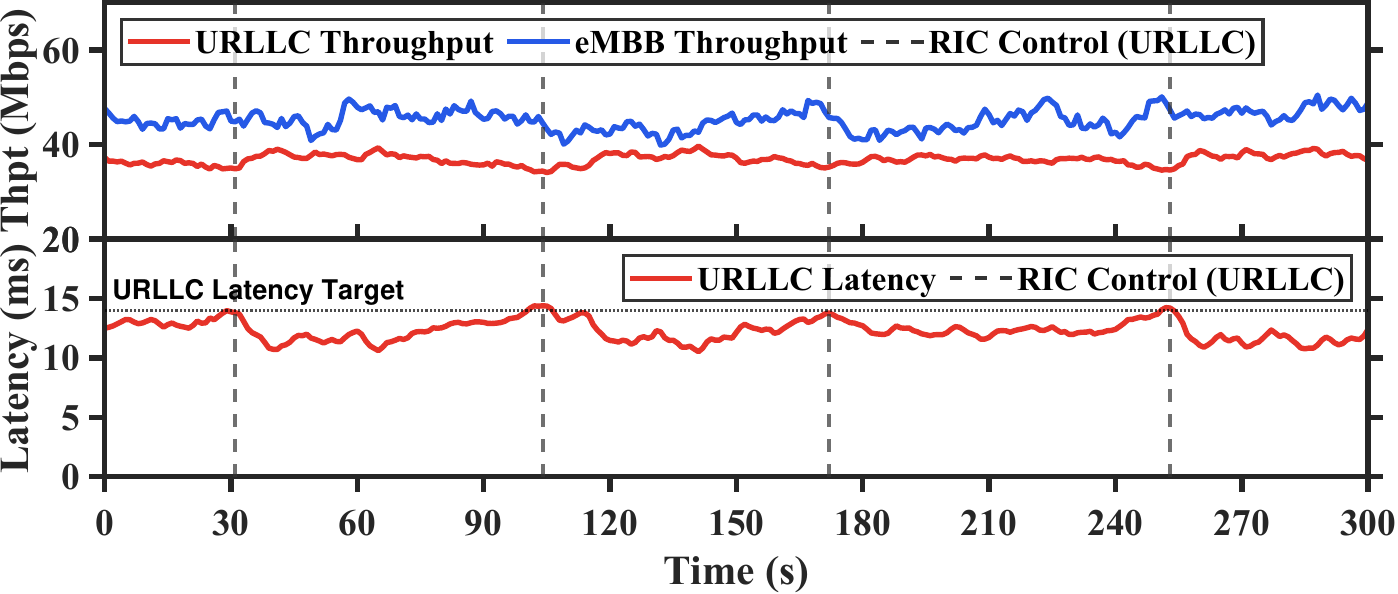}
    \caption{Runtime behavior of the generated slice-scheduling xApp}
    \label{fig:slice_scheduling_runtime}
    \vspace{-5pt}
\end{figure}

\textbf{Runtime Behavioral Check.}
\update{Beyond execution time, we further examine the runtime behavior of the generated xApps through two representative tasks. For anomaly detection, we use collected KPM traces with ground-truth anomalies to examine whether the xApp correctly detects anomalies while keeping false alarms and missed alarms at a low level. As shown in \figurename~\ref{fig:behavioral_test}, the generated xApp raises clear alarm outputs for most ground-truth anomalies, while the few false alarms and missed alarms are explicitly marked. For slice scheduling, we further evaluate the closed-loop control behavior of the generated xApp using live KPM measurements. Based on per-UE downlink throughput and configured slice requirements, the xApp dynamically determines PRB resource quotas and sends control requests to the E2 Node through E2SM-RC Style~2. As shown in \figurename~\ref{fig:slice_scheduling_runtime}, the vertical dashed lines indicate actual control requests, while the subsequent throughput and latency traces illustrate the resulting network behavior. These experiments demonstrate that the generated xApps can perform both runtime monitoring and network control.}


\subsection{Ablation Study}
\label{sec:ablation}
We conducted ablation studies to separately evaluate the contribution of each technical module. Specifically, we consider four key modules: \textit{Requirement Refinement and Structuring}, \textit{Background Knowledge Retrieval}, \textit{xApp Function Design}, and \textit{Code Validation}. In each ablation setting, we disabled one module while keeping the others intact, then generated 25 xApps per dataset and recorded two metrics: the one-shot success rate and the iteration-to-success count.

\begin{table}[t]
\centering
\setlength{\abovecaptionskip}{4pt}
\caption{Ablation study (\textbf{R}equirement refinement and structuring, \textbf{K}nowledge retrieval, \textbf{F}unction design, and \textbf{V}alidation.)}
\resizebox{0.47\textwidth}{!}{%
\begin{tabular}{c|cc|cc}
\bottomrule[2pt]
\multirow{2}{*}{\textbf{Method}} 
& \multicolumn{2}{c|}{\textbf{One-Shot Success Rate}} 
& \multicolumn{2}{c}{\textbf{Iteration-to-Success Count}}  \\
\cline{2-5}
& Baseline 1 & Baseline 2
& Baseline 1 & Baseline 2
\\
\hline 
w/o R. (\S~\ref{sec:user_interface}) & 0.88 & 0.84  & 8 & 3\\
w/o K. (\S~\ref{sec:knowledge})& 0.76 & 0.80 & 14 & 5 \\
w/o F. (\S~\ref{sec:function_design})& 0.90 & 0.80 & 8 & 3 \\
w/o V. (\S~\ref{sec:function_design})& 0.84 & 0.92  & 10 & 6\\
\hline
\textsc{\textbf{\projectName}} & \textbf{0.92} & \textbf{0.96} & \textbf{5} & \textbf{3}  \\
\toprule[2pt]
\end{tabular}
\label{tab:ablation}
}\vspace{-15pt}
\end{table}

The results shown in \tableautorefname{}~\ref{tab:ablation} demonstrate that the key modules collectively achieve the robust and reliable xApp synthesis with high performance and distinct contributions. The ablation of any single module results in a noticeable degradation in the overall performance. 
\update{Specifically, a significant drop occurs when the \textit{Background Knowledge Retrieval} module is excluded. Without this module, \projectName cannot retrieve and select the task-relevant O-RAN information needed during generation, such as KPM definitions, optimization heuristics, and algorithm references, and instead relies on the knowledge already available to the LLM. As a result, the generated xApps tend to adopt simplistic algorithms with generic logic flow, leading to a one-shot success rate below 80\%. We also observe a substantial increase in the number of code refinement iterations.}

In addition, the absence of the \textit{Requirement Refinement and Structuring} module also causes poor performance. Without it, user inputs are directly passed to LLMs in an ambiguous and unstructured form, hampering \projectName's ability to accurately interpret user intents. Consequently, the generated xApps often produce misaligned outputs with incorrect formats. Worse still, when the \textit{xApp Function Design} module is disabled, \projectName skips important intermediate reasoning steps without CoT prompting. This results in code that appears to be syntactically correct but often suffers from logical flaws and incomplete functionality implementations. 
Lastly, the ablation of \textit{Code Validation} module would disable the automated syntax checking and static analysis before final output, which leads to more immediate runtime failures.

In summary, these findings emphasize that the complete pipeline architecture of \projectName is essential to achieve consistent performance for generating functional xApps. Such a complementary combination of all technical modules allows \projectName to reliably generate executable xApps across diverse O-RAN control scenarios.

\vspace{-5pt}
\section{Related Work}
\label{sec:related_work}
\vspace{-2pt}

\noindent\textbf{xApp Development.} O-RAN promotes openness and modularity, enabling flexible integration of heterogeneous components across the RAN. Numerous studies~\cite {brik2024explainable,agarwal2025open,foukas2016flexran,foukas2017network,groen2024implementing,ko2024edgeric,calagna2025cormo,rachakonda2024comprehensive} have explored various aspects of O-RAN, but xApp development~\cite{santos2025managing} remains cumbersome due to system complexity and evolving specifications \cite{kilinc2022jade}. To facilitate deployment efficiency and multi-service support, OREO \cite{mungari2025ran} formulates xApp orchestration as a multi-service deployment problem and enables xApp sharing. Spotlight~\cite{sun2024spotlight} focuses on functionality design and proposes a telemetry-driven xApp for explainable anomaly detection. In control intelligence, DRL-based xApps~\cite{martinez2024drl} adopt deep reinforcement learning to allocate O-RAN resources and meet traffic and slicing demands. Recent work~\cite{wu2025llm} explores natural-language-based intent translation for O-RAN slice control, but it focuses on simple rule-based control and does not support code synthesis or validation. In contrast, \textit{\projectName presents an end-to-end automation framework for agile xApp development}, which allows developers or even network operators to quickly transform high-level business ideas and into readily deployable xApps and thereby drastically shorten the xApp development cycle.

\noindent\textbf{LLMs for Code Generation.}
By exploiting the remarkable language understanding and processing capabilities of LLMs, numerous intelligent applications have been developed, such as mobile task automation \cite{wen2024autodroid, lee2024mobilegpt} and IoT data interpretation \cite{xu2024penetrative, ji2024hargpt}. LLM-based code generation \cite{shen2025autoiot, shen2025gpiot, islam2024mapcoder, huang2023agentcoder} automatically translates natural-language user requirements into executable programs. A pioneering work MapCoder \cite{islam2024mapcoder} decomposes the code generation process into four stages---retrieval, planning, coding, and debugging---each orchestrated by LLM prompts. Recent AIoT-oriented systems, such as AutoIoT \cite{shen2025autoiot} and GPIoT \cite{shen2025gpiot}, further show that natural-language programming can be adapted to domain-specific IoT and AIoT application development through intent refinement, knowledge retrieval, code synthesis, and iterative validation.
\update{Recent studies have also explored LLMs specifically for O-RAN. ORANSight-2.0~\cite{gajjar2025oransight} demonstrates that O-RAN-specific model training can improve domain knowledge and code-related capabilities. AI5GTest~\cite{ganiyu2025ai5gtest} shows that explicit 3GPP and O-RAN specifications can be incorporated into LLM-based systems for automated testing and validation. These results reinforce two observations that motivate \projectName: O-RAN-specific knowledge is important when LLMs operate on O-RAN tasks, and specification-aware validation is necessary when generated outputs interact with O-RAN components.}

\update{Building on these observations and the natural-language programming paradigm demonstrated by prior AIoT systems, \projectName addresses the next step of automating xApp development. O-RAN xApps must interact with the near-RT RIC, subscribe to E2/KPM telemetry, satisfy A1 policy and E2 interface constraints, and meet near-RT control-loop requirements. \projectName therefore connects high-level operator requir ements with task- and platform-specific O-RAN knowledge retrieval, xApp function generation and validation, platform-specific interface synthesis, deployment, and runtime validation. This end-to-end process transforms the O-RAN knowledge and specification-aware capabilities demonstrated by recent LLM systems into a workflow for generating deployable near-RT RIC xApps.}

\vspace{-5pt}
\section{Conclusion}
\label{sec:conclusion}
\vspace{-2pt}
\projectName marks the paradigm shift in xApp development, from current manual programming by experts to fully automated xApp generation and deployment. 
\projectName is motivated to streamline xApp development with the latest advances in LLMs and agentic AI, and thus simplify xApp programming, shorten new feature launch time, and stimulate O-RAN innovation.  
To this end, \projectName builds an end-to-end automated xApp generation pipeline integrating a set of novel techniques for user intent elicitation, knowledge retrieval, code generation, and deployment. 
Evaluations show \projectName can generate 
performant xApps (on par with \update{manually developed} SOTA baselines) with little programming effort at a much shorter time. 
These encouraging results underscore the potential of \projectName in accelerating O-RAN innovation. 






\ifCLASSOPTIONcaptionsoff
  \newpage
\fi



%
\bibliographystyle{IEEEtran}
\bibliography{main}
\clearpage
%








\end{document}